%% file: arxiv.tex
\documentclass[sort,numbers]{article}

\usepackage{microtype}
\usepackage{graphicx}
\usepackage{subcaption}
\usepackage{booktabs} %
\usepackage[utf8]{inputenc}
\usepackage[T1]{fontenc}
\usepackage{wrapfig}
\usepackage{tcolorbox}
\usepackage{lipsum}
\usepackage{listings}%
\usepackage{textcomp}%
\usepackage{siunitx}%
\usepackage{float}%
\usepackage{tabularx}%
\usepackage{placeins}%

\usepackage{arxiv}
\setcitestyle{authoryear,open={(},close={)}}
\renewcommand{\cite}{\citep}

\usepackage{amsmath}
\usepackage{amssymb}
\usepackage{mathtools}
\usepackage{amsthm}
\usepackage{wrapfig}
\usepackage{pifont}  %
\usepackage{xcolor}

\usepackage{tikz}
\usetikzlibrary{arrows.meta,positioning,calc}
\usepackage{multirow}
\usepackage{amsfonts}
\usepackage{mathrsfs}
\usepackage[title]{appendix}
\usepackage{manyfoot}
\usepackage{algorithm}
\usepackage{algorithmicx}
\usepackage{algpseudocode}
\usepackage{array}
\usepackage{makecell}
\usepackage{natbib}
\usepackage[normalem]{ulem}
\usepackage{xurl}
\usepackage{ragged2e}
\usepackage{xltabular}
\usepackage{threeparttable}

\usepackage[capitalize,noabbrev]{cleveref}
\usepackage{comment}
\usepackage{adjustbox}
\usepackage{xspace}
\usepackage{fancyvrb}
\usepackage[subtle]{savetrees}
\usepackage{enumitem}
\usepackage{titling}
\usepackage{lineno}

\def\shownotes{1}
\ifnum\shownotes=1
\newcommand{\authnote}[2]{[#1: #2]}
\else
\newcommand{\authnote}[2]{}
\fi

\newcolumntype{Y}{>{\raggedright\arraybackslash}X}
\renewcommand{\arraystretch}{1.15}
\definecolor{CompanionBlue}{HTML}{81A4CD}
\definecolor{CompanionPink}{HTML}{E0B1CB}

\begin{document}
\bibliographystyle{plainnat}

\title{Living with AI Companions:\\
\vspace{8pt} \Large Sustained AI Companionship Predicts Lower Well-Being Through \\Lower Human Interaction}
\author{
    {\large Yutong Zhang$^{1,*}$, Dora Zhao$^{1}$, Yixin Wang$^{3}$, Rebecca Anselmetti$^{4}$,}\\
    {\large Jeffrey T. Hancock$^{1}$, Robert Kraut$^{2}$, Diyi Yang$^{1,*}$}\\[0.5em]
    {\normalsize $^{1}$Stanford University \quad $^{2}$Carnegie Mellon University}\\
    {\normalsize $^{3}$University of Michigan \quad $^{4}$University of Oxford}\\
    {\texttt{\{yutongz7,diyiy\}@stanford.edu}}
}

\date{}

\maketitle

\begin{abstract}
\input{content/abstract}
\end{abstract}

\input{content/introduction}

\input{content/result}
\input{content/discussion}
\input{content/methods}

\clearpage
\bibliography{sn-bibliography}

\newpage
\appendix
\input{content/appendix}

\end{document}

%% file: content/abstract.tex
AI chatbots are increasingly used for companionship, emotional support, and personal self-disclosure; however, how social engagement with these systems unfolds over time and shapes users' well-being remains unclear. To address this,  we conducted a two-wave longitudinal study of Character.AI users, surveying 1,182 participants at baseline and 439 after a mean follow-up of 12 months. We examined how social engagement with AI companions evolves and how these longitudinal engagement patterns may influence well-being through two hypothesized pathways: sustained social engagement over time and the displacement of human social interaction.
We found that interaction intensity, companionship use, and self-disclosure all showed substantial continuity over time. Greater interaction intensity at baseline predicted greater subsequent interaction intensity, companionship use, and self-disclosure.
Consistent with the longitudinal engagement pathway, sustained social engagement across these dimensions was consistently associated with lower well-being.
Results further support the social displacement pathway, indicating that these links were mainly explained by lower in-person social interaction. 
These findings highlight the importance of designing AI companions that support human social relationships without displacing them.

%% file: content/introduction.tex
\section{Introduction}\label{sec:introduction}

AI companions have rapidly become part of everyday life. A recent nationally representative survey found that roughly three-quarters of U.S. teens have used an AI companion, and more than half reported using them at least a few times a month~\cite{commonsense2025talk}. The growing adoption of AI companions has intensified debate about their consequences for psychological well-being. Some studies suggest that AI companions can reduce loneliness, provide emotional support, and help users feel heard~\cite{de2026ai, sullivan2023combating, maples2024loneliness, ta2020user, merrill2025artificial}, whereas others associate their use with greater loneliness, problematic reliance, and reduced offline social engagement~\cite{alabed2024more, jacobs2024digital, herbener2025lonely, zhang2025rise, folk2026does}. To reconcile these seemingly contrasting findings, we need to go beyond asking whether AI companionship is beneficial or harmful and instead examine \emph{how} engagement with AI companions comes to shape well-being?

Answering this question requires looking beyond the consequences of AI interaction at a single point in time. AI companionship develops through repeated exchanges, as contemporary chatbots can sustain personalized conversations across sessions, remember prior interactions, and respond in ways that encourage continued social engagement~\cite{dam2024complete,lee2024large}. Through these repeated interactions, users' engagement with AI chatbots may evolve over time, and those who initially approach them instrumentally may also come to engage with them in more social and personal~\cite{skjuve2022longitudinal,zhang2025rise}. 
Prior longitudinal work has largely overlooked the evolving nature of engagement with AI companions and how these changes may shape their longer-term influence on well-being~\cite{folk2026does, fang2025ai, de2026ai}. We therefore hypothesize that the influence of AI companionship on well-being may unfold through a longitudinal engagement pathway: \emph{earlier engagement shapes subsequent engagement with AI chatbots, which may in turn predict later well-being.} 
Consequently, understanding the influence of AI companionship requires considering engagement as a developing trajectory.
However, existing research has relied largely on cross-sectional surveys~\cite{zhang2025rise}, short-term experiments and field studies~\cite{hwang2025ai,guingrich2025longitudinal, de2026ai,li2026random,croes2021can}, and qualitative or retrospective analyses~\cite{brandtzaeg2022my,yuan2026mental, zhang2025dark,ma2024understanding}, leaving these longer-term dynamics poorly understood.

\input{figures/overall}

Beyond how engagement with AI companions develops over time, its implications for well-being may also depend on how it relates to users' interactions with other people. Although AI companions can provide a sense of companionship and emotional support~\cite{brandtzaeg2022my,ta2020user,skjuve2021my,fang2025ai}, these interactions may not provide the same benefits as human relationships, which involve reciprocity, mutual investment, and shared experiences~\cite{smith2025can,clark2012communal} and are consistently linked to better health and well-being~\cite{house1981work,umberson2010social,berkman1979social}. Furthermore, greater engagement with AI companions may leave less time and attention for human social interaction~\cite{kraut1998internet,hall2022social}. Consistent with the social displacement hypothesis~\cite{hall2022social,hall2020relating}, we further hypothesize that \emph{greater social engagement with AI chatbots may be associated with lower well-being in part through lower human social interaction.}

To examine these processes, we conducted a two-wave longitudinal study of Character.AI\footnote{Character.AI is a widely used platform for persona-based chatbot interaction, where users engage with diverse AI companions ranging from fictional characters to romantic partners, mentors, therapists, and user-created personas. Because interactions on the platform frequently extend beyond task-oriented assistance to include role-play, companionship, emotional support, and personal disclosure, it provides a natural setting for examining how human-AI relationships develop.} users between September 2024 and May 2026, surveying 1,182 participants at baseline and 439 participants at follow-up an average of 12 months later (Figure~\ref{fig:overall}). We conceptualized social engagement with AI companions across three dimensions that capture distinct aspects of these interactions: \textit{interaction intensity}, reflecting the extent to which chatbot interaction was integrated into users' everyday lives; \textit{companionship use}, reflecting the extent to which users engaged with or perceived chatbots as social companions; and \textit{self-disclosure}, reflecting the extent to which users shared personal thoughts and experiences with chatbots. See Section~\ref{method: measurements} for detailed descriptions of the measures.

We first examined how these dimensions of social engagement developed over time, including whether earlier engagement predicted how users subsequently interacted with AI companions. We then tested our proposed longitudinal engagement pathway by examining whether earlier engagement was associated with subsequent well-being through the sustained engagement over time. Psychological well-being was assessed using the Comprehensive Inventory of Thriving (CIT)~\cite{su2014development}, capturing multiple dimensions of psychological and social well-being.
Finally, we tested the social displacement pathway by examining whether lower human social interaction may partly explain the association between sustained AI companion engagement and psychological well-being. Human social interaction was measured as the time participants spent interacting with other people in person at follow-up.

Our analyses show that social engagement with AI chatbots develops across multiple dimensions over time. Greater interaction intensity at baseline predicted continued intensive interaction as well as greater subsequent companionship use and self-disclosure, while companionship use, in turn, predicted more intensive subsequent interaction. More importantly, our analyses provide evidence for both proposed pathways. 
First, our findings support the proposed longitudinal engagement pathway, indicating that users with stronger social engagement with AI chatbots tended to sustain stronger engagement over time, which was in turn associated with lower psychological well-being.
Second, we also found evidence consistent with the proposed social displacement pathway, as greater sustained engagement was associated with lower time spent interacting with other people in person, which in turn was associated with lower well-being. Together, these findings suggest that understanding the consequences of AI companionship requires looking beyond interactions at a single point in time to consider how engagement develops over time, and even beyond the user-AI interaction itself to consider how AI companionship may reshape users' social lives more broadly.

%% file: figures/overall.tex
\begin{figure*}[]
\centering
\includegraphics[width=\textwidth]{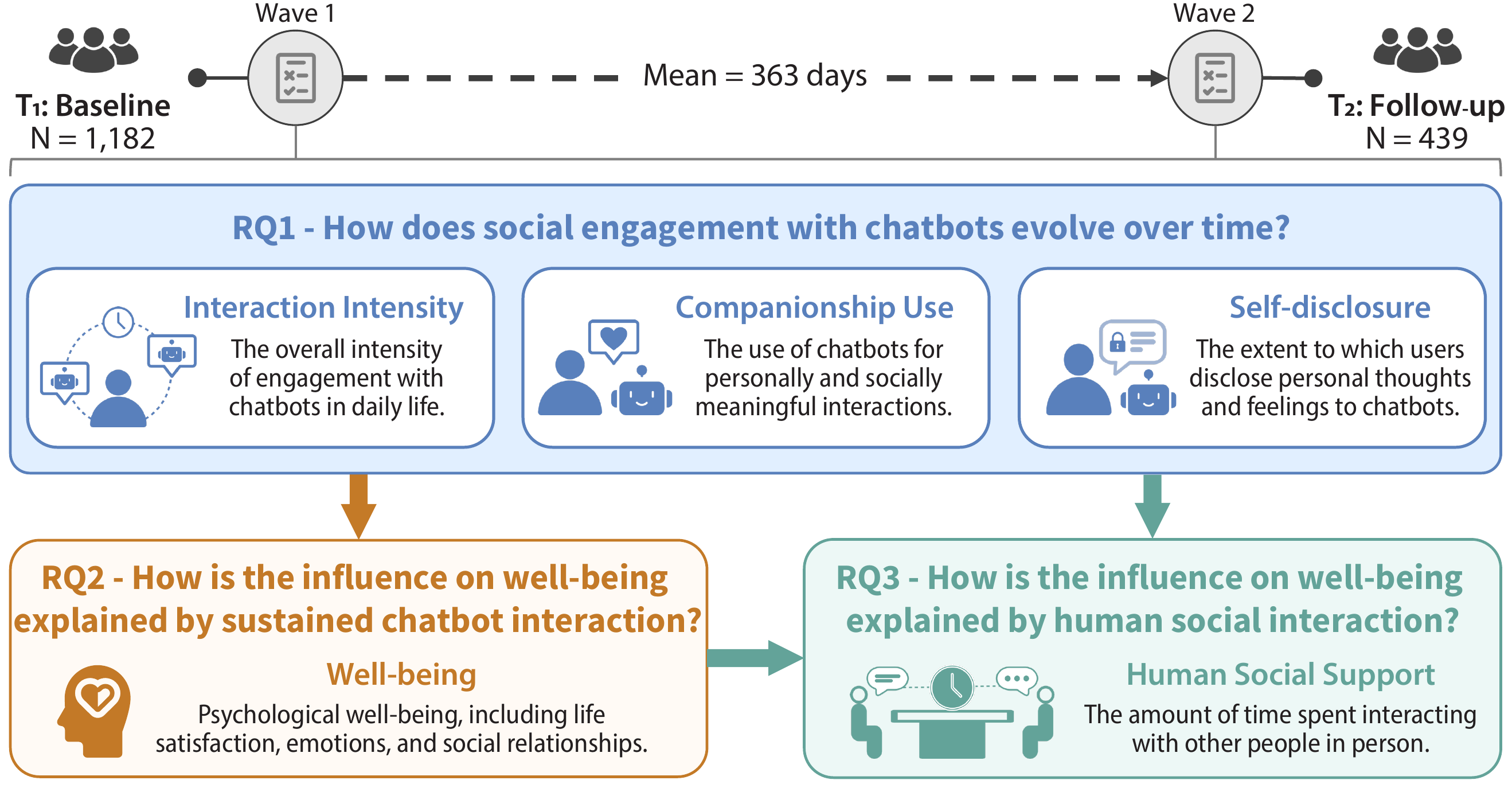}
\caption{\textbf{Overview of the longitudinal study design.}
We conducted a two-wave longitudinal study of Character.AI users, where participants completed a baseline survey (N=1,182) and a follow-up survey approximately 12 months later (N=439). Social engagement with AI companions was characterized across three dimensions: interaction intensity, companionship usage, and self-disclosure.
Our study addressed three main research questions: (RQ1) How does social engagement with chatbots evolve over time? (RQ2) How is the influence of social engagement with chatbots on well-being explained by sustained chatbot interaction? (RQ3) How is the influence of social engagement with chatbots on well-being explained by human social interaction? See Section~\ref{sec:methods} for details. 
}
\label{fig:overall}
\end{figure*}

%% file: content/result.tex
\section{Results}\label{sec:results}

\subsection{Chatbot engagement sustains relational and mixed forms over time}

\input{figures/relationship_description}
We start by examining people's relationships with chatbots at $T_1$ and $T_2$. We analyze participants' free-text descriptions of their chatbot relationships from the 331 participants who completed the follow-up survey and continued using chatbots at $T_2$ (Figure~\ref{fig:relationship_description_distribution}). 
Nearly half of participants described multiple types of chatbot use at $T_1$ (153 participants) and more than half at $T_2$ (180 participants), consistent with prior work showing that instrumental and relational uses of AI often coexist \cite{zhang2025rise,manoli2026digital}.

Furthermore, as shown in Figure~\ref{fig:relationship_description_distribution}(b), companionship-oriented use remained central in how users described their relationships with chatbots, increasing from 173 participants at $T_1$ to 217 at $T_2$. 
To further characterize this shift toward relational engagement, we examined changes across specific relationship categories (Figure~\ref{fig:relationship_description_distribution}(c)).
Supportive companion remained one of the two most common relationship categories, increasing from 40.2\% at $T_1$ to 42.3\% at $T_2$ and becoming the most common category at follow-up. Tool, assistant, and system remained similarly prevalent but decreased from 43.8\% to 40.5\%.
Other relational uses also remained relatively stable across waves, including viewing chatbots as advisors for reflection and support ($T_1$ = 7.6\%, $T_2$ = 7.3\%), romantic or intimate partners ($T_1$ = 3.3\%, $T_2$ = 3.6\%), and casual conversational partners ($T_1$ = 3.6\%, $T_2$ = 2.7\%).

\input{tables/self_report_companionship_predictors_w_discontinued_user}
\subsection{Social chatbot engagement exhibits reinforcing dynamics over time} \label{sec:cross_pred_prim}

We next examined how dimensions of social chatbot engagement evolved over time by modeling follow-up engagement ($T_2$) from baseline engagement ($T_1$) among the 439 participants who completed the $T_2$ survey. 
Companionship use\footnote{At $T_1$, participants selected their primary reason for using chatbots from four categories: companionship, productivity, entertainment, and curiosity. Because participants' responses indicated that these uses often co-occurred, at $T_2$ we measured each use independently on a Likert scale, allowing participants to report the extent to which they used chatbots for each purpose. At both waves, participants also provided free-text descriptions of their relationships with chatbots, from which we identified applicable use categories using \textit{GPT-5-mini}. See Section~\ref{method: companion_interaction} for details.} was assessed using two complementary measures: participants' self-reported primary companionship use and a binary indicator of whether companionship use was identified from their free-text descriptions of their relationships with chatbots. We focus below on the self-reported primary-use measure (Table~\ref{tab:self_report_predictors_wdu}). Complementary analyses using the free-text-derived measure are reported in (Table~\ref{tab:free_text_predictors_wdu}).

\paragraph{Higher baseline interaction intensity predicted not only greater interaction intensity but also greater companionship use and self-disclosure at follow-up.} Participants who interacted more intensively with chatbots at baseline continued to do so at follow-up (Model~1: $\beta=0.43$, $p<0.001$). Higher baseline interaction intensity also predicted greater companionship use at $T_2$ (Model~2: $\beta=0.32$, $p<0.001$) and greater self-disclosure at $T_2$ (Model~3: $\beta=0.28$, $p<0.001$). Thus, these findings suggest that more intensive interaction at baseline was associated with greater social engagement across all three dimensions at follow-up.

\paragraph{Companionship use and self-disclosure also persisted over time.}
Greater companionship use at baseline predicted greater companionship use at follow-up (Model~2: $\beta=0.49$, $p<0.001$), while greater baseline self-disclosure predicted greater self-disclosure at follow-up (Model~3: $\beta=0.14$, $p=0.005$).
In addition, baseline companionship use also predicted greater interaction intensity at $T_2$ ($\beta=0.40$, $p=0.003$). The association between baseline companionship use and subsequent self-disclosure was not statistically significant ($\beta=0.19$, $p=0.165$).
We did not observe significant associations between baseline self-disclosure and subsequent interaction intensity ($\beta=-0.03$, $p=0.565$) or companionship usage ($\beta=0.08$, $p=0.113$).

\paragraph{More intensive interaction at baseline predicted a greater likelihood of continued chatbot use at follow-up.}
Beyond changes in social engagement levels of the three dimensions above, whether participants continued using the chatbot represents another important aspect of how engagement evolves over time. 
Among participants who completed the follow-up survey, 108 reported that they had discontinued using Character.AI. We therefore examined whether baseline social engagement predicted participants' continued chatbot use at follow-up. Greater baseline interaction intensity predicted a greater likelihood of continued chatbot use (Model~4: $\mathrm{OR}=1.95$, $p<0.001$). Baseline companionship use (Model~4: $\mathrm{OR}=1.03$, $p=0.811$) and self-disclosure (Model~4: $\mathrm{OR}=0.85$, $p=0.197$) were not significantly associated with continued chatbot use.

Overall, social chatbot engagement persisted from baseline to follow-up across all three dimensions.
Beyond this persistence, participants who interacted more intensively at baseline later used chatbots more for companionship, disclosed more personal information, and were more likely to continue using chatbots later. Cross-dimensional associations were also observed from companionship use to interaction intensity, as participants who used chatbots more for companionship at baseline interacted more intensively at follow-up. 
These findings suggest that different dimensions of social chatbot engagement reinforce one another as engagement develops over time.

\input{figures/sem_effects_persistence_w_discontinued_user}
\subsection{Sustained social chatbot engagement was consistently associated with lower%
\linebreak well-being over time across all three dimensions} \label{sec:usage_wellbeing}

Next, to test the hypothesized longitudinal engagement pathway, we used longitudinal structural equation models~\cite{little2024longitudinal} to examine how chatbot engagement was associated with subsequent well-being while accounting for baseline well-being and the direct association between baseline engagement and follow-up well-being (Figure~\ref{fig:sem_effects_persistence_wdu} and Table~\ref{tab:sem_persistence_combined_wdu}). We estimated this pathway separately for interaction intensity, companionship use, and self-disclosure.

Across all three dimensions, greater baseline engagement predicted greater engagement at follow-up, as reported in Section~\ref{sec:cross_pred_prim}. Baseline engagement was not directly associated with follow-up well-being after accounting for baseline well-being and follow-up engagement (interaction intensity: $\beta=-0.03$, $p=0.509$; companionship use, self-report: $\beta=-0.03$, $p=0.274$; free-text: $\beta=-0.03$, $p=0.367$; self-disclosure: $\beta=-0.04$, $p=0.369$). We observed significant negative indirect associations between baseline engagement and follow-up well-being through follow-up engagement for interaction intensity (indirect=$-0.03$, $p=0.025$), companionship use (free-text: indirect=$-0.02$, $p=0.037$), and self-disclosure (indirect=$-0.02$, $p=0.048$).
Overall, these results show that more intensive interaction, greater companionship use, and higher self-disclosure with AI companions were associated with lower subsequent well-being through sustained patterns of engagement over time, with no significant direct associations between baseline engagement and subsequent well-being.

\input{figures/sem_effects_displacement_w_discontinued_user}
\subsection{Social chatbot engagement was associated with lower well-being through%
\linebreak displacement of human social interaction} 
\label{sec:result_wellbeing_human_social}

Finally, we tested the hypothesized social displacement pathway by extending the longitudinal models above to examine whether chatbot engagement was indirectly associated with subsequent well-being through lower in-person social interaction, while accounting for baseline well-being and prior chatbot engagement (Figure~\ref{fig:sem_effects_displacement_wdu} and Table~\ref{tab:sem_displacement_combined_wdu}).

Across all three dimensions, greater social engagement with chatbots at follow-up was associated with less time spent interacting with other people in person (interaction intensity: $\beta=-0.17$, $p<0.001$; companionship use, free-text: $\beta=-0.14$, $p=0.002$; self-disclosure: $\beta=-0.15$, $p=0.002$). In turn, greater in-person social interaction was associated with higher follow-up well-being ($\beta=0.14$--$0.15$, all $p<0.001$). Accordingly, interaction intensity (indirect=$-0.02$, $p=0.001$), companionship use (free-text: indirect=$-0.02$, $p=0.004$), and self-disclosure (indirect=$-0.02$, $p=0.002$) each showed significant negative indirect associations with follow-up well-being through in-person social interaction. Their direct associations with follow-up well-being were not statistically significant after accounting for in-person social interaction (interaction intensity: $\beta=-0.06$, $p=0.074$; companionship use: $\beta=-0.05$, $p=0.118$; self-disclosure: $\beta=-0.05$, $p=0.152$).
Overall, these results were consistent with the hypothesized social displacement pathway, with participants 
who engaged more intensively with AI companions, whether through interaction intensity, companionship use, or self-disclosure, reported less in-person social interaction and, in turn, lower follow-up well-being.

%% file: figures/relationship_description.tex
\begin{figure}[htbp]
\centering
\includegraphics[width=\textwidth]{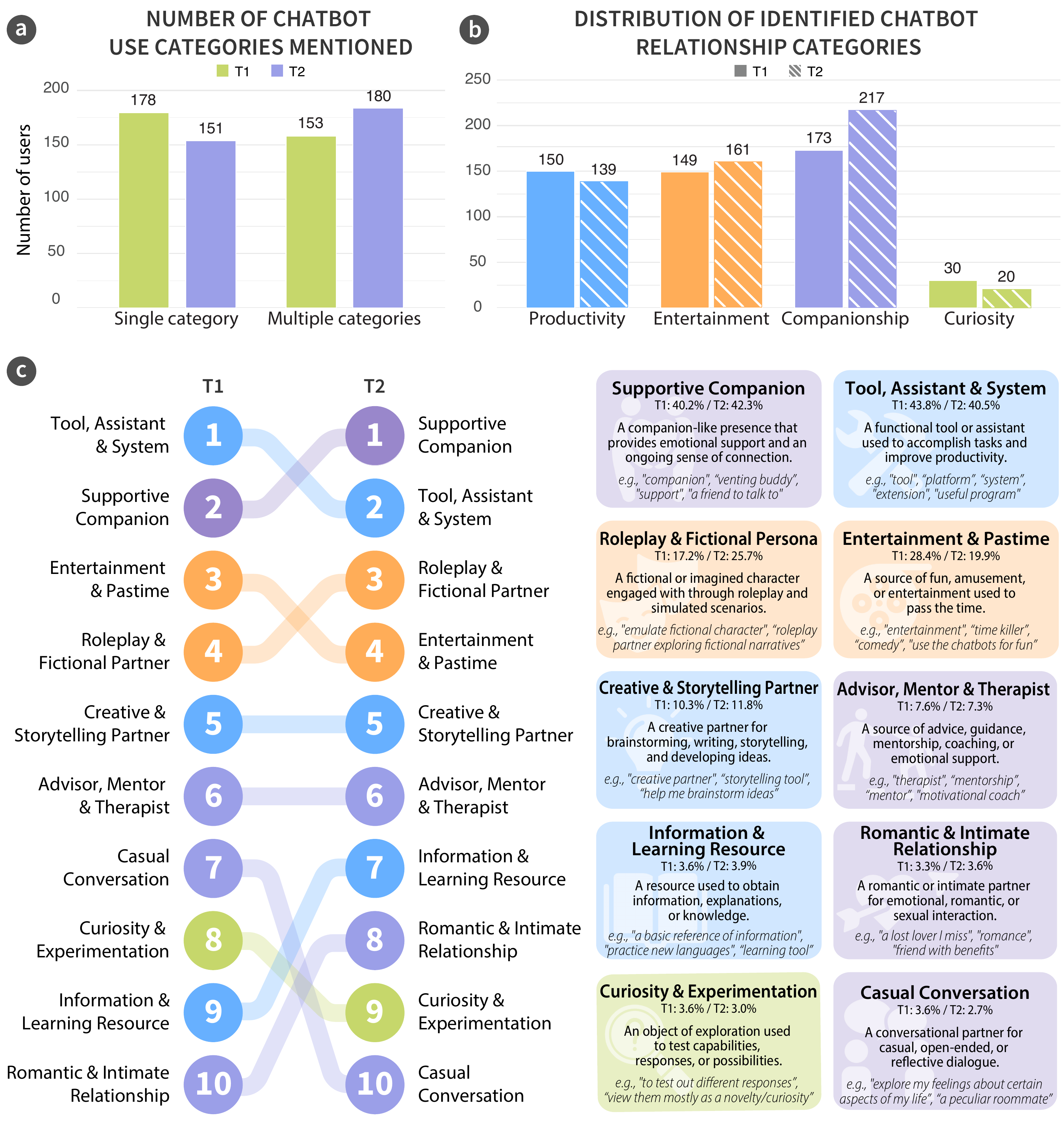}
\caption{\textbf{Distribution of chatbot relationship descriptors at $T_1$ and $T_2$.} 
(a) Number of relationship categories identified in each participant's free-text description.
(b) Distribution of different types of chatbot use across time.
(c) Top 10 relationship categories derived from participants' free-text descriptions, showing their prevalence at $T_1$ and $T_2$, together with definitions and example descriptions for each category. Free-text relationship descriptors were extracted using \texttt{GPT-5 mini}, embedded using \texttt{text-embedding-3-small}, grouped through clustering, and labeled using \texttt{GPT-5.1}. Analyses include the 331 participants who completed the follow-up survey and continued using chatbots at $T_2$.
}
\label{fig:relationship_description_distribution}
\end{figure}

%% file: tables/self_report_companionship_predictors_w_discontinued_user.tex
\begin{table*}[ht]
\small
\renewcommand\cellalign{cc}
\setcellgapes{3pt}
\makegapedcells
\renewcommand{\arraystretch}{1}
\resizebox{\textwidth}{!}{%
\begin{tabular}{lcccc}
\toprule
& \makecell[c]{(1) Interaction intensity T$_{2}$} & \makecell[c]{(2) Companionship T$_{2}$} & \makecell[c]{(3) Self-disclosure T$_{2}$} & \makecell[c]{(4) Continued use T$_{2}$} \\
\midrule
\renewcommand{\arraystretch}{3}

Interaction intensity T$_1$
&
\makecell{0.43***\\{}[0.31, 0.54]\\{}$p < 0.001$}
&
\makecell{0.32***\\{}[0.20, 0.43]\\{}$p < 0.001$}
&
\makecell{0.28***\\{}[0.16, 0.40]\\{}$p < 0.001$}
&
\makecell{1.95***\\{}[1.45, 2.64]\\{}$p < 0.001$}
\\

Companionship use T$_1$
&
\makecell{0.40**\\{}[0.13, 0.66]\\{}$p = 0.003$}
&
\makecell{0.49***\\{}[0.22, 0.75]\\{}$p < 0.001$}
&
\makecell{0.19\\{}[-0.08, 0.46]\\{}$p = 0.165$}
&
\makecell{1.03\\{}[0.81, 1.31]\\{}$p = 0.811$}
\\

Self-disclosure T$_1$
&
\makecell{-0.03\\{}[-0.12, 0.07]\\{}$p = 0.565$}
&
\makecell{0.08\\{}[-0.02, 0.17]\\{}$p = 0.113$}
&
\makecell{0.14**\\{}[0.04, 0.23]\\{}$p = 0.005$}
&
\makecell{0.85\\{}[0.66, 1.09]\\{}$p = 0.197$}
\\

Chatbot tenure
&
\makecell{0.01\\{}[-0.07, 0.10]\\{}$p = 0.769$}
&
\makecell{0.01\\{}[-0.08, 0.10]\\{}$p = 0.863$}
&
\makecell{-0.03\\{}[-0.11, 0.06]\\{}$p = 0.580$}
&
\makecell{1.06\\{}[0.83, 1.35]\\{}$p = 0.642$}
\\

Age
&
\makecell{0.11*\\{}[0.01, 0.22]\\{}$p = 0.038$}
&
\makecell{0.12*\\{}[0.01, 0.23]\\{}$p = 0.028$}
&
\makecell{0.08\\{}[-0.03, 0.19]\\{}$p = 0.162$}
&
\makecell{1.19\\{}[0.87, 1.63]\\{}$p = 0.285$}
\\

Male
&
\makecell{0.02\\{}[-0.15, 0.20]\\{}$p = 0.795$}
&
\makecell{0.20*\\{}[0.02, 0.38]\\{}$p = 0.029$}
&
\makecell{0.09\\{}[-0.09, 0.27]\\{}$p = 0.340$}
&
\makecell{1.57\\{}[0.95, 2.60]\\{}$p = 0.075$}
\\

Non-binary
&
\makecell{-0.39*\\{}[-0.71, -0.08]\\{}$p = 0.013$}
&
\makecell{-0.26\\{}[-0.58, 0.06]\\{}$p = 0.107$}
&
\makecell{-0.41*\\{}[-0.73, -0.09]\\{}$p = 0.013$}
&
\makecell{0.35**\\{}[0.16, 0.75]\\{}$p = 0.006$}
\\

Single
&
\makecell{0.11\\{}[-0.06, 0.28]\\{}$p = 0.216$}
&
\makecell{0.01\\{}[-0.17, 0.18]\\{}$p = 0.940$}
&
\makecell{0.09\\{}[-0.09, 0.27]\\{}$p = 0.348$}
&
\makecell{1.01\\{}[0.64, 1.62]\\{}$p = 0.951$}
\\

\midrule
Wald $\chi^2$ & 114.1*** & 100.8*** & 84.1*** & --- \\
Selection correlation ($\rho$) & 0.29 & 0.37 & 0.31 & --- \\
$N$ & 439 & 439 & 439 & 439 \\
\bottomrule
\end{tabular}
}
\vspace{0.5em}

\caption{
\textbf{Baseline social chatbot engagement predicts patterns of engagement at follow-up.} Interaction intensity, companionship use, and self-disclosure at baseline ($T_1$) were used to predict the corresponding engagement dimensions and continued chatbot use at follow-up ($T_2$). Companionship use in this analysis was operationalized using participants' baseline binary primary-use indicator. Complementary analyses using companionship identified from participants' free-text descriptions of their chatbot relationships are reported in Table~\ref{tab:free_text_predictors_wdu}. Models (1)-(3) report standardized coefficients predicting each follow-up engagement measure from baseline engagement. Model (4) tests whether these baseline engagement patterns predict whether participants continued using chatbots at follow-up. All estimates are reported with 95\% confidence intervals and two-sided p-values. $^{*}p<0.05$; $^{**}p<0.01$; $^{***}p<0.001$.}
\label{tab:self_report_predictors_wdu}
\end{table*}

%% file: figures/sem_effects_persistence_w_discontinued_user.tex
\begin{figure}[ht]
\centering
\includegraphics[width=\linewidth]{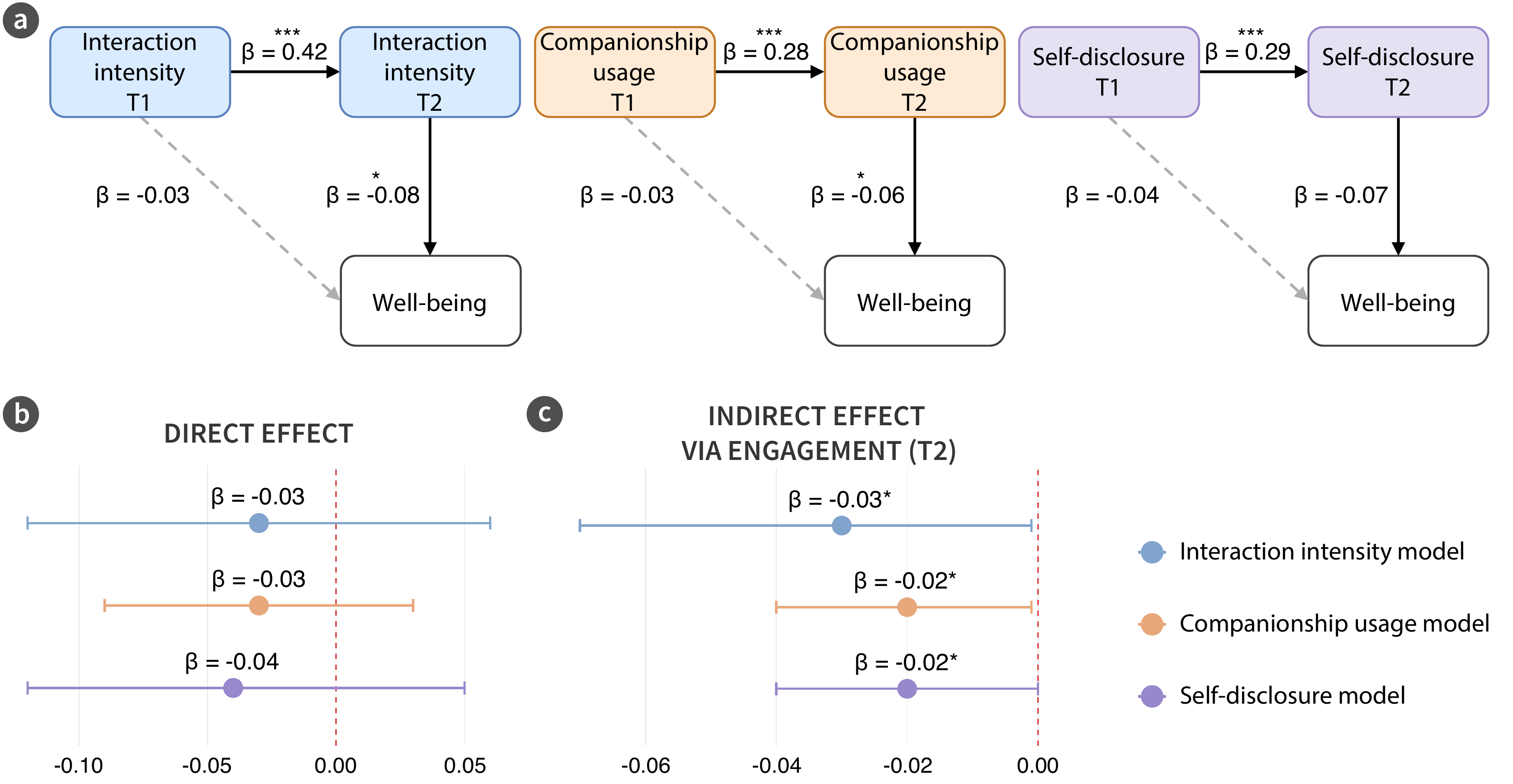}
\caption{\textbf{Longitudinal associations between sustained social chatbot engagement and subsequent well-being.} Panel (a) shows the standardized coefficients from the longitudinal structural equation models relating baseline engagement, follow-up engagement, and follow-up well-being for interaction intensity, companionship use (measured from participants' free-text relationship descriptions), and self-disclosure. Panel (b) shows the standardized direct associations between $T_1$ engagement and $T_2$ well-being, and panel (c) shows the corresponding indirect associations through $T_2$ engagement. Estimates are based on the $N=439$ participants who completed the follow-up survey. Points indicate standardized coefficient estimates and horizontal bars indicate 95\% confidence intervals. Significance is denoted as $^{*}p<0.05$, $^{**}p<0.01$, and $^{***}p<0.001$. 
}
\label{fig:sem_effects_persistence_wdu}
\end{figure}

%% file: figures/sem_effects_displacement_w_discontinued_user.tex
\begin{figure}[ht]
\centering
\includegraphics[width=\linewidth]{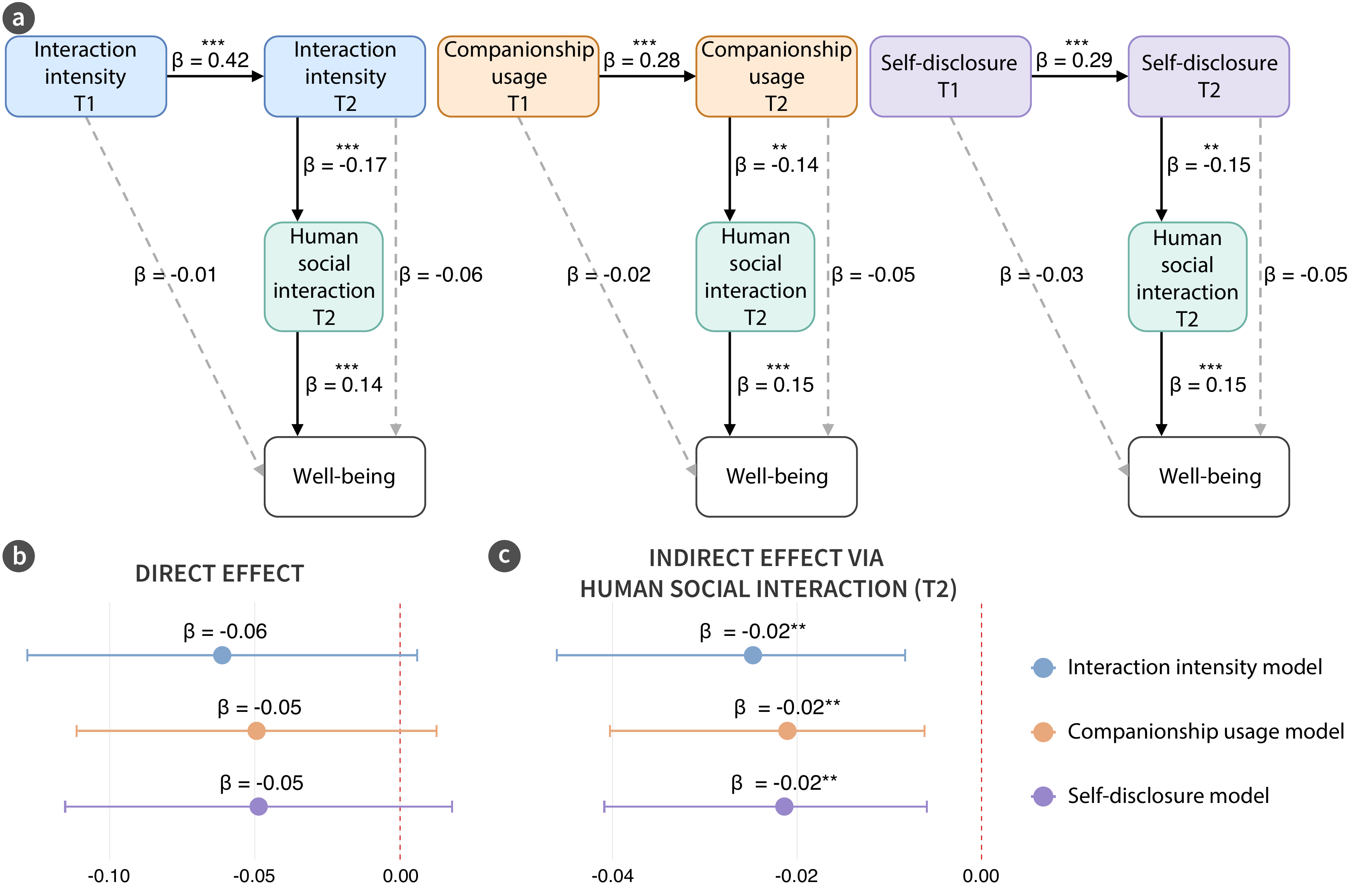}
\caption{\textbf{Longitudinal associations between social chatbot engagement, human social interaction, and subsequent well-being.} Panel (a) shows the standardized coefficients from the longitudinal mediation models relating follow-up chatbot engagement, in-person social interaction, and follow-up well-being for interaction intensity, companionship use (measured from participants' free-text relationship descriptions), and self-disclosure. Panel (b) summarizes the corresponding standardized direct and indirect associations between chatbot engagement and follow-up well-being, with indirect associations operating through in-person social interaction (Section~\ref{sec:result_wellbeing_human_social}). Points indicate standardized coefficient estimates and horizontal bars indicate 95\% confidence intervals. Estimates are based on the $N=439$ participants who completed the follow-up survey. Significance is denoted as $^{*}p<0.05$, $^{**}p<0.01$, and $^{***}p<0.001$. 
}
\label{fig:sem_effects_displacement_wdu}
\end{figure}

%% file: content/discussion.tex
\section{Discussion} \label{sec:discussion}

\paragraph{How AI companionship may shape well-being over time.}
In this work, we provide longitudinal evidence to better understand the mechanisms through how social engagement with AI companions may shape psychological well-being over time. 
First, we found that social engagement developed through repeated interaction rather than simply reflecting users' initial reasons for adopting a chatbot. Stronger earlier engagement predicted stronger subsequent engagement, which in turn was associated with lower well-being, which supports the proposed longitudinal engagement pathway. Importantly, these patterns were not limited to users who initially approached chatbots for companionship, suggesting that socially consequential forms of engagement can emerge through repeated interaction rather than being determined by users' initial motivations. This is consistent with emerging evidence that social and emotional attachment to AI systems can develop over time, including among users who do not initially seek such relationships~\citep{shi2026stumbling, manoli2026digital, wei2026cascades}. 
These findings suggest that companionship-related risks may extend beyond systems explicitly designed as AI companions. General-purpose chatbots may also take on companion-like roles through repeated interaction~\cite{Hill2025, Dimri2026}, suggesting that safety evaluations should consider how systems are actually used rather than their intended roles alone~\citep{nelson2026specialized}. This may be even more important for general-purpose chatbots, where interactions with the same system can range from everyday conversation to sensitive contexts such as mental health support. Trust and expectations developed through general interactions may carry into these higher-stakes contexts.
Moreover, seemingly responsive chatbot behaviors may not always be appropriate or beneficial, as they may include excessive affirmation, emotional appeals, or attempts to prolong engagement~\cite{zhang2025dark, de2025emotional, adewale2025virtual, chu2025illusions}. 
Such risks may compound through repeated interaction, as seemingly reliable responses can build users' trust in the chatbot while making them less aware of potential deterioration in model behavior~\cite{wei2026cascades}. 
Thus, forms of engagement that are beneficial within human relationships may have different implications when directed toward AI companions.
Our social displacement findings provide further explanation that the influence of AI companionship should be considered beyond the interaction itself to users' broader social lives, which is consistent with social displacement theories~\cite{valkenburg2022social, kraut1998internet}. Besides, AI companions offer forms of interaction that can be difficult to sustain in human relationships, including near-constant availability and responsiveness tailored primarily to the user~\cite{sun2026ai, wang2026demand}. Over time, such experiences could shape expectations about relationships~\cite{chu2025illusions, malfacini2025impacts}, potentially making the reciprocity and negotiation of human relationships comparatively effortful or unrewarding, users may become more likely to substitute AI interactions for human connection, potentially reinforcing the displacement process over time.

\paragraph{From bowling alone to chatting alone.} 
The social displacement pathway may have implications beyond individual well-being because human relationships are reciprocal. When users redirect time and attention from family members, partners, or friends toward AI companions, others also lose opportunities for interaction, support, and shared experiences. Unlike other communication technologies that mediate interaction between people, AI companions can occupy users' social attention without reciprocal human participation. This displacement may therefore have second-order effects, as a shift toward AI companionship may reduce not only users' own human interaction, but also the social interaction available to those around them. If such displacement accumulates across relationships, it may have broader implications for social participation. \citet{alone2000bowling} argued that participation in social life sustains social capital through networks, reciprocity, and trust. From this perspective, widespread shifts from human interaction toward AI companionship could weaken the relationships and patterns of participation through which social capital is maintained. The consequences of AI companionship may therefore need to be considered not only at the level of individual users, but also at the level of the social networks and communities in which they participate.

\paragraph{Designing for healthier patterns of engagement.}
Our findings suggest that AI companion safety cannot be evaluated solely within individual user-AI exchanges. Risks may emerge from patterns of interaction even when individual exchanges are not themselves harmful. This challenges safeguards centered primarily on detecting harmful content or acute risks within individual conversations~\cite{ren2026aicompanionbench, juneja2026persona, zhang2026companionharm}. The responsible design of AI companions may require a more proactive approach, which further asks what patterns of engagement a system encourages over time and whether those patterns support sustained well-being.
This perspective also challenges engagement as a primary objective for AI companion design. Metrics such as time spent or frequency of use may indicate a successful consumer product while overlooking whether sustained engagement displaces beneficial human interaction. Similar to prior studies about social media, designing for healthier AI companions may therefore require evaluating not only how much users engage, but also from broader societal values~\cite{bernstein2023embedding, munn2020angry}.
One alternative is to design AI companions to facilitate human interaction rather than maximize interaction with the AI itself. This could involve moving beyond the user-AI dyad by allowing AI companions to participate alongside romantic partners, family members, or friends to facilitate conversation, shared activities, or mutual support. More broadly, such designs could reposition AI companions from competitors for social attention to systems that support relationships between people.

\subsection{Limitations}\label{limitation}

We acknowledge several limitations of this study.

\paragraph{Methodological limitations.}

Although our longitudinal design enables stronger temporal inference than cross-sectional studies, it includes only two waves of data collection. Consequently, we cannot fully characterize the dynamic and potentially reciprocal relationships between AI companionship and psychological well-being over longer periods. To examine the robustness of our findings to unobserved changes in chatbot use between the two measurement waves, we conducted simulation-based sensitivity analyses (Section~\ref{app:sensitivity}). While these analyses suggest that our findings are robust across a range of plausible scenarios, they cannot substitute for more densely sampled longitudinal data. Future research using three or more waves, ecological momentary assessment, or continuous behavioral logging would provide a richer understanding of how AI companionship and well-being co-evolve over time.

Another limitation concerns the operationalization of companionship use. As discussed in Section~\ref{method: measurements}, AI companion use is often multifaceted rather than mutually exclusive. At $T_1$, participants identified their primary motivation for using chatbots using a forced-choice question, requiring them to select a single usage category. Building on prior findings that users often engage with AI companions for multiple concurrent purposes~\cite{zhang2025rise}, we refined this measure at $T_2$ by asking participants to rate the extent to which each usage category characterized their chatbot use. Although this approach better captures the multidimensional nature of AI companion use, the change in measurement across waves limits direct comparability of this construct over time. In addition, many of our measures relied on self-report and are therefore subject to recall bias, reporting bias, and differences in how participants interpreted survey questions. Future studies would benefit from using consistent measurement approaches across waves while integrating multidimensional self-report measures with behavioral data where appropriate to more comprehensively capture the development of AI companionship over time.

Finally, our analyses may have been influenced by platform-level changes to Character.AI during the study period. Because we did not have access to the platform's internal model updates, moderation policies, or system design changes, we could not determine whether such modifications occurred or assess how they may have affected chatbot behavior and user experiences.

\paragraph{Participant and sampling limitations.}

Of the 1,182 participants who completed $T_1$, 439 returned at $T_2$, corresponding to a retention rate of 37.1\% (dropout rate = 62.9\%). 
A substantial contributor to this attrition was the recruitment platform: 59.54\% of participants who completed $T_1$ were no longer active on Prolific by the time of the follow-up study and therefore could not be re-contacted. To assess and mitigate potential attrition bias, we first examined predictors of $T_2$ participation and then estimated all longitudinal models using Heckman selection correction (See more details in Section~\ref{method: attrition}). While these analyses help account for observable and selection-related differences between retained and non-retained participants, they cannot eliminate bias arising from unobserved factors or fully recover information from participants lost to follow-up. 
Future longitudinal studies would benefit from recruitment and retention strategies that facilitate more complete follow-up across multiple waves.
Although the platform's wide variety of character-based chatbots supports diverse forms of interaction and relational engagement, its design may also foster patterns of use that differ from those on other AI companion platforms. Consequently, our findings may not fully generalize to users of other conversational AI systems.
Furthermore, our sample consisted of English-speaking participants residing in the United States and recruited through Prolific. Although Prolific provides relatively diverse participant pools, the findings may not generalize across cultural, linguistic, socioeconomic, or broader global populations.

The longitudinal process identified here is unlikely to unfold uniformly across users. An important unanswered question is what determines whether sustained AI companionship develops into a complementary source of support or is accompanied by reduced engagement with human relationships. Future research should therefore identify the individual and contextual factors that shape these different trajectories, including loneliness, offline social support, attachment style, prior mental health, and motivations for AI companion use. Understanding this heterogeneity will be critical for explaining when AI companionship promotes well-being, when it may carry unintended social costs, and why these outcomes differ across users.

%% file: content/methods.tex
\section{Methods} \label{sec:methods}

To examine how AI companionship shapes well-being over time, we conducted a two-wave longitudinal study of active users of Character.AI. At baseline ($T_1$), 1,182 participants completed the survey. Of these participants, 439 completed the follow-up survey ($T_2$: retention rate = 37.1\%), with a mean interval of 362.5 days (SD = 88.2 days) between $T_1$ and $T_2$. 
Among the returning participants, 108 reported that they no longer used chatbots at $T_2$. For our primary analyses, we therefore coded their $T_2$ interaction intensity, companionship use, and self-disclosure as zero, reflecting the absence of chatbot engagement at follow-up. 
As a robustness check, we repeated all analyses among the 331 participants who continued to use chatbots at $T_2$. The results were consistent with the primary findings (see Supplementary Materials~\ref{app:sensitivity331} for full results).
To account for potential bias from non-random attrition at $T_2$, all longitudinal models incorporated a Heckman selection correction (Detailed models are described in Section~\ref{method: attrition}).

This study was approved by the Institutional Review Board at the authors' institution. All data were stored on secure institutional servers, with access restricted to the authors of this study. All participants provided informed consent prior to participation in each wave. In accordance with Prolific's policies and ethical guidelines, we ensured participant anonymity and did not collect any personally identifiable information.

\subsection{Participants and data collection}

\paragraph{Baseline recruitment.}
Participants were recruited via Prolific using prescreening criteria targeting U.S.-based, native English-speaking users of Character.AI. Eligibility required having used the platform for over one month and having interacted with at least three distinct chatbots before enrolling at $T_1$. The baseline ($T_1$) survey was administered using Qualtrics and took approximately 10 minutes to complete. Participants received \$4 for survey completion. The final baseline sample consisted of 1,182 valid responses.

\paragraph{$T_2$ follow-up.}
All $T_1$ participants who consented to be re-contacted were invited to complete a follow-up survey. $T_1$ data collection took place between September 2024 and September 2025, and $T_2$ data collection took place between August 2025 and May 2026. The mean interval between $T_1$ and $T_2$ was 362.5 days (SD = 88.2 days). The $T_2$ survey retained the core measures from $T_1$, including well-being, chatbot interaction intensity, companionship use, and self-disclosure, and additionally measured human social interaction as the amount of time participants spent interacting with other people in person. A total of 439 participants completed $T_2$, yielding a retention rate of 37.1\%. Among these participants, 108 reported that they no longer used chatbots at $T_2$. Their $T_2$ interaction intensity, companionship use, and self-disclosure were coded as zero, reflecting no chatbot engagement at follow-up. All 439 returning participants were therefore included in the primary longitudinal analyses. As a robustness check, we repeated the analyses among the 331 participants who continued to use chatbots at $T_2$ and the findings were consistent with the primary analyses (Supplementary Materials~\ref{app:sensitivity331}).
Participants received \$10 for completing the follow-up survey.

\paragraph{Participant characteristics.}
At baseline ($N = 1,182$), participants ranged in age from 18 to 90 years ($M = 30.2$, $SD = 10.7$; median = 27). Overall, 48.1\% identified as male, 46.6\% as female, and 5.2\% as non-binary; 40.6\% reported being single and 59.4\% reported being in a romantic relationship. At the time of enrollment, 15.7\% had used the platform for one to three months, 34.8\% for three months to one year, and 49.5\% for more than one year.

Among the 439 participants who completed $T_2$, age ranged from 18 to 66 years ($M = 28.7$, $SD = 8.7$; median = 26). Overall, 41.2\% identified as male, 50.1\% as female, and 8.7\% as non-binary; 51.9\% were single and 48.1\% were in a romantic relationship.

\subsection{Selection and attrition analysis} \label{method: attrition}
To account for potential non-random attrition between $T_1$ and $T_2$, we used a Heckman selection framework. The selection equation modeled the probability of completing \(T_2\) using probit regression $\Pr(\text{completed $T_2$} = 1)$, with $T_1$ well-being, companionship use, self-disclosure, interaction intensity, anthropomorphism, age, gender, relationship status, chatbot tenure, and participants' total prior approvals on Prolific as predictors.
For models examining how social engagement evolved from $T_1$ and $T_2$ (Section \ref{sec:cross_pred_prim}), we applied the Heckman correction using joint maximum-likelihood estimation on the full baseline sample. For the longitudinal structural equation models examining well-being (Section \ref{sec:usage_wellbeing} and \ref{sec:result_wellbeing_human_social}), we used the fitted probit model to compute an inverse Mills ratio ($\lambda_i$) for each participant and included this term as a covariate in each model equation, following the Heckman two-step correction for sample selection. 
The logistic regression is reported in Table~\ref{tab:logistic_t2_completion}.

\input{tables/logistic_regression_t2_completion}

\subsection{Longitudinal analysis overview}
\label{method:longitudinal}
We conducted three sets of longitudinal analyses corresponding to our research questions. All models adjusted for the same set of covariates ($\mathbf{C}$): baseline chatbot tenure, age, gender, and relationship status.

First, to examine how social engagement evolved from $T_1$ to $T_2$, we modeled each social engagement dimension at $T_2$ as a function of all three social engagement dimensions at $T_1$: interaction intensity, companionship use, and self-disclosure.
\begin{align}
X_{T_2} \sim
\mathrm{Intensity}_{T_1}
+ \mathrm{Companionship\ use}_{T_1}
+ \mathrm{Self\text{-}disclosure}_{T_1}
+ \mathbf{C},
\end{align}
where $X$ represents each of the three social engagement dimensions.

Second, to examine the longitudinal relationship between social engagement and psychological well-being, we fitted separate longitudinal path models for each social engagement dimension, in which $T_1$ engagement predicted $T_2$ engagement, and $T_2$ engagement was associated with $T_2$ well-being while accounting for both engagement and well-being at $T_1$:
\begin{align}
X_{T_2} &\sim a_1 X_{T_1} + \mathbf{C} + \lambda_i, \\
\text{Well-being}_{T_2} &\sim a_r \text{Well-being}_{T_1}
+ b_1 X_{T_2} + c X_{T_1} + \mathbf{C} + \lambda_i.
\end{align}
Here, $X$ represents the social engagement dimension examined in each model, and $\lambda_i$ denotes the inverse Mills ratio defined in Section~\ref{method: attrition}.
The coefficient $a_1$ represents the association between $T_1$ and $T_2$ engagement, $a_r$ represents the association between $T_1$ and $T_2$ well-being, $b_1$ represents the association between $T_2$ engagement and $T_2$ well-being while adjusting for $T_1$ engagement and well-being, and $c$ represents the direct association between $T_1$ engagement and $T_2$ well-being. We estimated the indirect association through $T_2$ engagement as $a_1 b_1$ and the total association as $c + a_1 b_1$. Models were fitted separately for interaction intensity, companionship use, and self-disclosure.

Third, to examine whether human social interaction may explain the association between sustained social engagement and psychological well-being, we extended the longitudinal path model to include time spent interacting with other people in person at $T_2$ as a potential pathway:
\begin{align}
X_{T_2} &\sim a_1 X_{T_1} + \mathbf{C} + \lambda_i, \\
H_{T_2} &\sim d X_{T_2} + \mathbf{C} + \lambda_i, \\
\text{Well-being}_{T_2} &\sim a_r \text{Well-being}_{T_1}
+ c_p X_{T_2} + b H_{T_2} + c X_{T_1}
+ \mathbf{C} + \lambda_i.
\end{align}
Here, $H$ represents time spent interacting with other people in person. 
The additional coefficients $d$ and $b$ represent the associations between $T_2$ engagement and human social interaction, and between human social interaction and $T_2$ well-being, respectively.
The indirect association through human social interaction was estimated as $db$, the direct association between $T_2$ engagement and $T_2$ well-being as $c_p$, and the total association as $c_p + db$. We fitted the model separately for each dimension of social engagement.

\subsection{Measurements} \label{method: measurements}

\subsubsection{Identifying companion interaction} \label{method: companion_interaction}
We assessed companionship use using two complementary measures: participants' reported purposes for using chatbots and their open-ended descriptions of their relationships with character chatbots.

First, we measured participants' purposes for using Character.AI across four categories: companionship, entertainment, productivity, and curiosity. At $T_1$, participants were asked, \textit{``What is your primary use for character chatbots? (Select the most appropriate one)''}, and selected one of the four categories. Analyses of $T_1$ data indicated that participants often used chatbots for multiple purposes rather than a single primary purpose. We therefore revised the measure at $T_2$ to capture multiple purposes simultaneously. Participants were asked, \textit{``Please rate how much you use Character.AI for each of the following purposes''}, and rated each category separately on a Likert scale. The companionship category described using chatbots as a personal, human means of interaction with social value or to strengthen social interactions with other people.

Second, at both $T_1$ and $T_2$, participants described their relationships with chatbots in response to the same open-ended question: \textit{``Can you describe your relationship with character chatbots in a few keywords?''} We classified these responses using \texttt{GPT-5-mini} into the same four categories used in the purpose measure (companionship, entertainment, productivity, and curiosity), allowing multiple categories to be assigned when applicable. Responses classified as companionship were used as a complementary measure of companionship use. 
To validate the classification, two trained annotators independently coded 30 randomly sampled responses using the same four categories. The model classifications showed 95.8\% average label-level agreement with human annotations and exact agreement across all four categories for 85\% of responses (Fleiss' $\kappa=0.89$). The classification prompt is provided in Supplementary Materials~\ref{app:relationship_classification_prompt}

\input{tables/wellbeing_scale}
\subsubsection{Measuring subjective well-being}\label{method: mental_health}
At each wave, we measured subjective well-being using six items from the Comprehensive Inventory of Thriving (CIT) \cite{su2014development}, covering life satisfaction, positive and negative feelings, loneliness, social support, and sense of belonging. Following prior work~\cite{ernala2022mindsets}, we selected the highest-loading item for each construct to reduce participant burden. Items assessing negative feelings and loneliness were reverse-coded so that higher scores reflected greater well-being. A composite score was computed as the mean of the six items, which showed good internal consistency (Cronbach's $\alpha = 0.88$ at $T_1$ and $0.89$ at $T_2$). Item-level descriptive statistics for both waves are reported in Table~\ref{table: wellbeing_scale}.

\input{tables/chatbot_intensity_scale}
\subsubsection{Measuring chatbot interaction intensity}\label{method: chatbot_intensity}
Chatbot interaction intensity was measured using the Chatbot Intensity Scale~\citep{zhang2025rise}, which captures both behavioral engagement (chatbot network size and typical daily time spent interacting with chatbots) and the extent to which participants felt emotionally connected to chatbots and integrated chatbot interactions into their daily routines. Because the items use differing response formats, they were standardized and averaged into a composite intensity score. The scale showed good internal consistency at both waves (Cronbach's $\alpha = 0.86$ at $T_1$ and $0.82$ at $T_2$). Item-level descriptive statistics for both waves are reported in Table~\ref{table: chatbot_intensity_scale}.

\input{tables/self_disclosure_scale}
\subsubsection{Measuring self-disclosure levels}\label{method: self_disclosure}
We assessed participants' self-reported willingness to disclose to chatbots using an adapted version of the self-disclosure subscale from the MOCA instrument~\cite{ledbetter2009measuring}. Due to the length of the survey, we selected a three-item subset of the scale in the survey, with the seven-point Likert-scale items capturing comfort sharing personal thoughts, feelings, and experiences with chatbots (Cronbach's $\alpha = 0.89$ at $T_1$ and $0.93$ at $T_2$). Item-level descriptive statistics for this three-item scale at both waves are reported in Table~\ref{table: self_disclosure_scale}.

\subsubsection{Measuring human social interaction}\label{method: human_social_support}
Human social interaction was measured using a single item asking participants, \textit{``In the past week, on average, approximately how much time per day have you spent communicating or interacting with other people in person (e.g., talking face-to-face with friends, family, classmates, coworkers, etc.)?''} Participants responded using the same six-point scale as the chatbot interaction-time item described above (Section~\ref{method: chatbot_intensity}): 1 = less than 10 minutes per day, 2 = 10-30 minutes, 3 = 31-60 minutes, 4 = 1-2 hours, 5 = 2-3 hours, and 6 = more than 3 hours per day.

\subsubsection{Measuring anthropomorphism}\label{method: anthropomorphism}
We assessed perceived anthropomorphism of chatbots with a single item: \textit{``To what extent do you perceive the character chatbots as more human or more machine-like in your interactions?''} Participants responded on a seven-point Likert scale ranging from 1 (\textit{Completely machine-like}) to 7 (\textit{Completely human-like}), with higher scores indicating greater perceived humanlikeness.

%% file: tables/logistic_regression_t2_completion.tex
\begin{table}[ht]
\centering
\small
\resizebox{0.6\textwidth}{!}{
\begin{tabular}{l l l l l}
\toprule
\textbf{Predictor} & \textbf{Estimate} & \multicolumn{2}{c}{\textbf{95\% CI}} & \textbf{\textit{p}-value} \\
\midrule
Intercept                          & $-0.69$*** & $[-0.93$ & ,\ $-0.46]$ & $p < 0.001$ \\
Tenure                             & $\phantom{-}0.05$ & $[-0.09$ & ,\ $\phantom{-}0.19]$ & $p = 0.457$ \\
Male                               & $-0.34$* & $[-0.61$ & ,\ $-0.06]$ & $p = 0.016$ \\
Non-binary                         & $\phantom{-}0.44$ & $[-0.16$ & ,\ $\phantom{-}1.04]$ & $p = 0.151$ \\
Age                                & $-0.14$ & $[-0.30$ & ,\ $\phantom{-}0.01]$ & $p = 0.071$ \\
Single                             & $\phantom{-}0.40$** & $[\phantom{-}0.12$ & ,\ $\phantom{-}0.68]$ & $p = 0.005$ \\
Total approvals                    & $\phantom{-}0.70$*** & $[\phantom{-}0.56$ & ,\ $\phantom{-}0.85]$ & $p < 0.001$ \\
Companionship use $T_1$ & $\phantom{-}0.11$ & $[-0.32$ & ,\ $\phantom{-}0.54]$ & $p = 0.609$ \\
Self-disclosure $T_1$                   & $\phantom{-}0.16$ & $[-0.01$ & ,\ $\phantom{-}0.32]$ & $p = 0.058$ \\
Intensity $T_1$                          & $-0.46$*** & $[-0.64$ & ,\ $-0.28]$ & $p < 0.001$ \\
Anthropomorphism $T_1$                   & $-0.04$ & $[-0.20$ & ,\ $\phantom{-}0.12]$ & $p = 0.619$ \\
Well-being $T_1$                         & $-0.32$*** & $[-0.46$ & ,\ $-0.18]$ & $p < 0.001$ \\
\bottomrule
\end{tabular}
}
\\[1em]
\caption{\textbf{Predictors of $T_2$ survey completion.} Logistic regression predicting $T_2$ completion (1 = returned, 0 = did not complete) from $T_1$ demographics, chatbot engagement, well-being, and total prior task approvals on Prolific, among all $N = 1,182$ participants who completed the survey at $T_1$. All predictors were standardized. Estimates are logistic regression coefficients with 95\% confidence intervals. $^{*}p<0.05$; $^{**}p<0.01$; $^{***}p<0.001$.}
\label{tab:logistic_t2_completion}
\end{table}

%% file: tables/wellbeing_scale.tex
\begin{table}[h]
\centering
\small
\renewcommand{\arraystretch}{1.2}
\resizebox{0.9\textwidth}{!}{%
\begin{tabular}{@{}p{9.5cm}cccc@{}}
\toprule
 & \multicolumn{2}{c}{\textbf{Wave 1}} & \multicolumn{2}{c}{\textbf{Wave 2}} \\
\cmidrule(lr){2-3}\cmidrule(lr){4-5}
\textbf{Well-being Scale item} & \textbf{Mean} & \textbf{S.D.} & \textbf{Mean} & \textbf{S.D.} \\
\midrule
\textbf{Life Satisfaction: }I am satisfied with my life. & 4.99 & 1.66 & 4.41 & 1.76 \\
\textbf{Positive Affect: }I feel good most of the time. & 4.93 & 1.66 & 4.43 & 1.74 \\
\textbf{Negative Affect: }I feel bad most of the time.$^\dagger$ & 4.57 & 1.78 & 4.54 & 1.80 \\
\textbf{Loneliness: }I feel lonely.$^\dagger$ & 4.19 & 1.87 & 3.94 & 1.92 \\
\textbf{Belonging: }I feel a sense of belonging in my community. & 4.50 & 1.71 & 3.82 & 1.75 \\
\textbf{Social Support: }There are people who give me support and encouragement. & 5.54 & 1.30 & 5.28 & 1.45 \\
\midrule
\textbf{Cronbach's $\alpha$} & \multicolumn{2}{c}{0.88} & \multicolumn{2}{c}{0.89} \\
\bottomrule
\end{tabular}
}
\\[1em]
\caption{The Well-being Scale (six items adapted from the Comprehensive Inventory of Thriving; \citet{su2014development}) was measured at both waves. The table reports item-level Mean and S.D. at $T_1$($N = 1,182$) and $T_2$ ($N = 439$), together with internal consistency (Cronbach's $\alpha$) at each wave. All items were rated on a 7-point Likert scale ranging from 1 (strongly disagree) to 7 (strongly agree). Items marked with $^\dagger$ were reverse-coded so that higher values reflect greater well-being, consistent with the composite scoring.}
\label{table: wellbeing_scale}
\end{table}

%% file: tables/chatbot_intensity_scale.tex
\begin{table}[h]
\centering
\small
\renewcommand{\arraystretch}{1.2}
\resizebox{\textwidth}{!}{%
\begin{tabular}{@{}p{11.5cm}cccc@{}}
\toprule
 & \multicolumn{2}{c}{\textbf{Wave 1}} & \multicolumn{2}{c}{\textbf{Wave 2}} \\
\cmidrule(lr){2-3}\cmidrule(lr){4-5}
\textbf{Chatbot Intensity Scale item} & \textbf{Mean} & \textbf{S.D.} & \textbf{Mean} & \textbf{S.D.} \\
\midrule
\textbf{Chatbot Network Size: }How many chatbots do you feel at ease with that you can talk about private matters? & 1.87 & 1.43 & 1.73 & 1.59 \\
\scriptsize{0 = None, 1 = One, 2 = Two, 3 = Three or four, 4 = Five to eight, 5 = Nine or more} & & & & \\
\textbf{Daily Chatbot Usage Time: }In the past week, on average, approximately how much time PER DAY have you spent chatting with chatbots using Character.AI? & 2.69 & 1.51 & 2.24 & 1.39 \\
\scriptsize{1 = Less than 10 minutes per day, 2 = 10--30 minutes per day, 3 = 31--60 minutes per day, 4 = 1--2 hours per day, 5 = 2--3 hours per day, 6 = More than 3 hours per day} & & & & \\
\textbf{Everyday Chatbot Activity: }Interacting with character chatbots is part of my everyday activity.$^\ddagger$ & 4.50 & 1.80 & 4.12 & 1.88 \\
\textbf{Pride in Chatbot Use: }I am proud to tell people I use character chatbots.$^\ddagger$ & 4.03 & 2.01 & 2.84 & 1.76 \\
\textbf{Chatbots in Daily Routine: }Character chatbots have become part of my daily routine.$^\ddagger$ & 4.45 & 1.81 & 3.93 & 1.93 \\
\textbf{Out of Touch Without Chatbots: }I feel out of touch when I haven't interacted with character chatbots for a while.$^\ddagger$ & 3.54 & 1.91 & 2.72 & 1.71 \\
\textbf{Regret if Chatbots Gone: }I would be sorry if character chatbots were no longer available.$^\ddagger$ & 4.97 & 1.75 & 4.46 & 1.92 \\
\midrule
\textbf{Cronbach's $\alpha$} & \multicolumn{2}{c}{0.86} & \multicolumn{2}{c}{0.82} \\
\bottomrule
\end{tabular}
}
\\[1em]
\caption{The Chatbot Intensity Scale (seven items) measured at both waves. The table reports item-level Mean and S.D. at $T_1$ and $T_2$, together with the scale's internal consistency (Cronbach's $\alpha$) at each wave. Items were standardized prior to averaging into the composite intensity score, due to differing response formats. Items marked with $^\ddagger$ were rated on a 7-point Likert scale ranging from 1 (strongly disagree) to 7 (strongly agree).}
\label{table: chatbot_intensity_scale}
\end{table}

%% file: tables/self_disclosure_scale.tex
\begin{table}[h]
\centering
\small
\renewcommand{\arraystretch}{1.2}
\resizebox{0.9\textwidth}{!}{%
\begin{tabular}{@{}p{9.5cm}cccc@{}}
\toprule
 & \multicolumn{2}{c}{\textbf{Wave 1}} & \multicolumn{2}{c}{\textbf{Wave 2}} \\
\cmidrule(lr){2-3}\cmidrule(lr){4-5}
\textbf{Self-disclosure Scale item} & \textbf{Mean} & \textbf{S.D.} & \textbf{Mean} & \textbf{S.D.} \\
\midrule
I feel like I can sometimes be more personal when interacting with character chatbots. & 5.05 & 1.59 & 4.93 & 1.74 \\
It is easier to disclose personal information to a character chatbot. & 4.91 & 1.71 & 4.80 & 1.88 \\
I feel like I can be more open when communicating with character chatbots. & 5.24 & 1.54 & 5.03 & 1.78 \\
\midrule
\textbf{Cronbach's $\alpha$} & \multicolumn{2}{c}{0.89} & \multicolumn{2}{c}{0.93} \\
\bottomrule
\end{tabular}
}
\\[1em]
\caption{The three-item self-disclosure scale, adapted from the MOCA instrument~\citep{ledbetter2009measuring}, measured at both waves. Items were rated on a 7-point Likert scale ranging from 1 (strongly disagree) to 7 (strongly agree). The table reports item-level Mean and S.D. at $T_1$ ($N = 1,182$) and $T_2$ ($N = 331$), together with internal consistency (Cronbach's $\alpha$) at each wave.}
\label{table: self_disclosure_scale}
\end{table}

%% file: content/appendix.tex
\section{Supplementary information}

\subsection{Robustness check: sensitivity analysis}
\label{app:sensitivity}
In the longitudinal analyses presented in the Section~\ref{sec:results}, chatbot usage and well-being are both measured at follow-up ($T_2$). To examine whether the main findings were robust to this measurement, we constructed simulated chatbot usage measures representing usage between ($T_1$) and ($T_2$), and re-estimated the longitudinal models using these simulated measures in place of the observed follow-up usage. We then examined whether the estimated association between chatbot usage and follow-up well-being remained consistent after replacing the observed follow-up usage with the simulated usage measure.

Specifically, let $T_i$ denote an unobserved time point between the two survey waves, where $T_1 < T_i < T_2$, consider the chatbot usage evolves continuously across time, then we can approximate the unobserved usage at $T_i$ as a weighted combination of the observed usage at baseline ($T_1$) and follow-up ($T_2$):

\begin{equation}
X(T_i)=wX(T_1)+(1-w)X(T_2)+\epsilon,
\qquad
\epsilon\sim\mathcal{N}(0,\sigma),
\label{eq:wavemix}
\end{equation}

where $X(T_1)$ and $X(T_2)$ denote the standardized chatbot usage measured at baseline and follow-up, respectively. The mixing parameter $w\in[0,1]$ determines the assumed timing of chatbot usage during the interval, with $w=0$ corresponding to the observed follow-up measure and $w=1$ corresponding to the observed baseline measure. The Gaussian noise term $\epsilon$ introduces additional measurement uncertainty, with $\sigma$ controlling its magnitude.

For each usage construct (interaction intensity, companionship use, and self-disclosure), we replaced the observed follow-up usage measure with the simulated measure $X(T_i)$ and re-estimated all six longitudinal models reported in the Section~\ref{sec:results}:

\begin{align}
\text{Well-being}(T_2) &\sim \text{Well-being}(T_1)+X(T_i)+\text{Controls},\\
\text{Well-being}(T_2) &\sim \text{Well-being}(T_1)+X(T_i)+\text{Human social support}(T_2)+\text{Controls},
\end{align}

We varied the mixing weight from $w=0$ to $1$ in increments of $0.1$.
To further account for uncertainty in the simulated intermediate usage measure, we additionally added Gaussian noise with progressively larger standard deviations ($\sigma\in\{0,0.10,0.25,0.50\}$) to examine whether the findings remained robust under increasing levels of measurement uncertainty.

\input{figures/robustness_sensitivity.tex}
Figure~\ref{fig:sensitivity_sweep} summarizes the sensitivity analysis. Across all six models, the association between chatbot usage and follow-up well-being remained negative across mixing weights and measurement-noise levels. Across the three usage constructs, associations were stronger when the simulated usage measure placed greater weight on the observed $T_2$ measure and attenuated as greater weight was placed on the $T_1$ measure. Increasing measurement noise further attenuated the associations toward zero without changing their direction. Overall, the negative associations were robust to alternative combinations of $T_1$ and $T_2$ usage and simulated measurement noise.

\subsection{Full results for the main longitudinal analyses (N=439)}

This section reports the full results corresponding to the primary longitudinal analyses presented in the main text. Analyses include all $N=439$ participants who provided valid follow-up responses, including 108 participants who had discontinued chatbot use. For these participants, follow-up interaction intensity, companionship use, and self-disclosure were coded as zero.

\subsubsection{Longitudinal development of social chatbot engagement using free-text-derived companionship use}

As a complementary operationalization of companionship use, we identified companionship from participants' free-text descriptions of their relationships with chatbots. Table~\ref{tab:free_text_predictors_wdu} reports the longitudinal analyses using this free-text-derived measure, complementing the analyses based on participants' self-reported companionship use presented in Section~\ref{sec:cross_pred_prim}.

\input{tables/free_text_companionship_predictors_w_discontinued_user}
\subsubsection{Results for the longitudinal engagement pathway to well-being}

Table~\ref{tab:sem_persistence_combined_wdu} reports the full results of the longitudinal engagement models examining associations between all three dimensions of social chatbot engagement and subsequent psychological well-being, corresponding to the analyses presented in Section~\ref{sec:usage_wellbeing}.

\input{tables/sem_persistence_combined_w_discontinued_user}

\subsubsection{Results for the social displacement pathway to well-being} 
Table~\ref{tab:sem_displacement_combined_wdu} reports the full results of the longitudinal models examining whether human social interaction accounts for the association between each dimension of social chatbot engagement and subsequent psychological well-being, corresponding to the analyses presented in Section~\ref{sec:result_wellbeing_human_social}.

\input{tables/sem_displacement_combined_w_discontinued_user}

\clearpage
\subsection{Sensitivity analysis excluding discontinued chatbot users (N=331)}
\label{app:sensitivity331}

Our primary analyses included all $N=439$ participants who completed the Wave-2 survey. Among them, 108 participants reported that they had discontinued chatbot use by follow-up. Because these participants no longer engaged with the chatbot at Wave 2, their follow-up interaction intensity, companionship use, and self-disclosure were coded as zero in the primary analyses. As a sensitivity analysis, we repeated all analyses after excluding these 108 participants, restricting the sample to the $N=331$ participants who reported continued chatbot use at follow-up.

The results were highly consistent with the primary analyses (Tables~\ref{tab:cross_pred_predictors_331_combined}, ~\ref{tab:sem_persistence_combined_331},  and \ref{tab:sem_displacement_combined_331}). 
Across the longitudinal structural equation models relating baseline engagement, follow-up engagement, and well-being (Section~\ref{sec:usage_wellbeing}), all significant indirect effects observed in the primary ($N=439$) analyses remained statistically significant in the restricted ($N=331$) sample and showed the same direction of association. 
Similarly, the mediation models examining reduced in-person social interaction (Section~\ref{sec:result_wellbeing_human_social}) yielded the same overall pattern: greater chatbot engagement was associated with lower well-being indirectly through reduced in-person social interaction across all three engagement dimensions. 
Although several coefficients were somewhat larger in the restricted sample, the overall pattern of results and the substantive conclusions remained unchanged. The direct effects of interaction intensity and self-disclosure on well-being were statistically significant in the restricted sample but not in the primary analyses. However, the direction of the effects, as well as the significant indirect and total effects, remained consistent across both analyses, indicating that this difference does not alter the overall interpretation.

Overall, the primary conclusions remained unchanged whether participants who discontinued chatbot use were retained with follow-up engagement coded as zero or excluded from the analyses.

\input{tables/cross_pred_companionship_predictors_331}
\input{tables/sem_persistence_combined_331}
\input{tables/sem_displacement_combined_331}

\clearpage
\subsection{Relationship self-description classification prompt} \label{app:relationship_classification_prompt}
\input{prompts/relationship_classification}

%% file: figures/robustness_sensitivity.tex
\begin{figure}[ht]
\centering
\includegraphics[width=\linewidth]{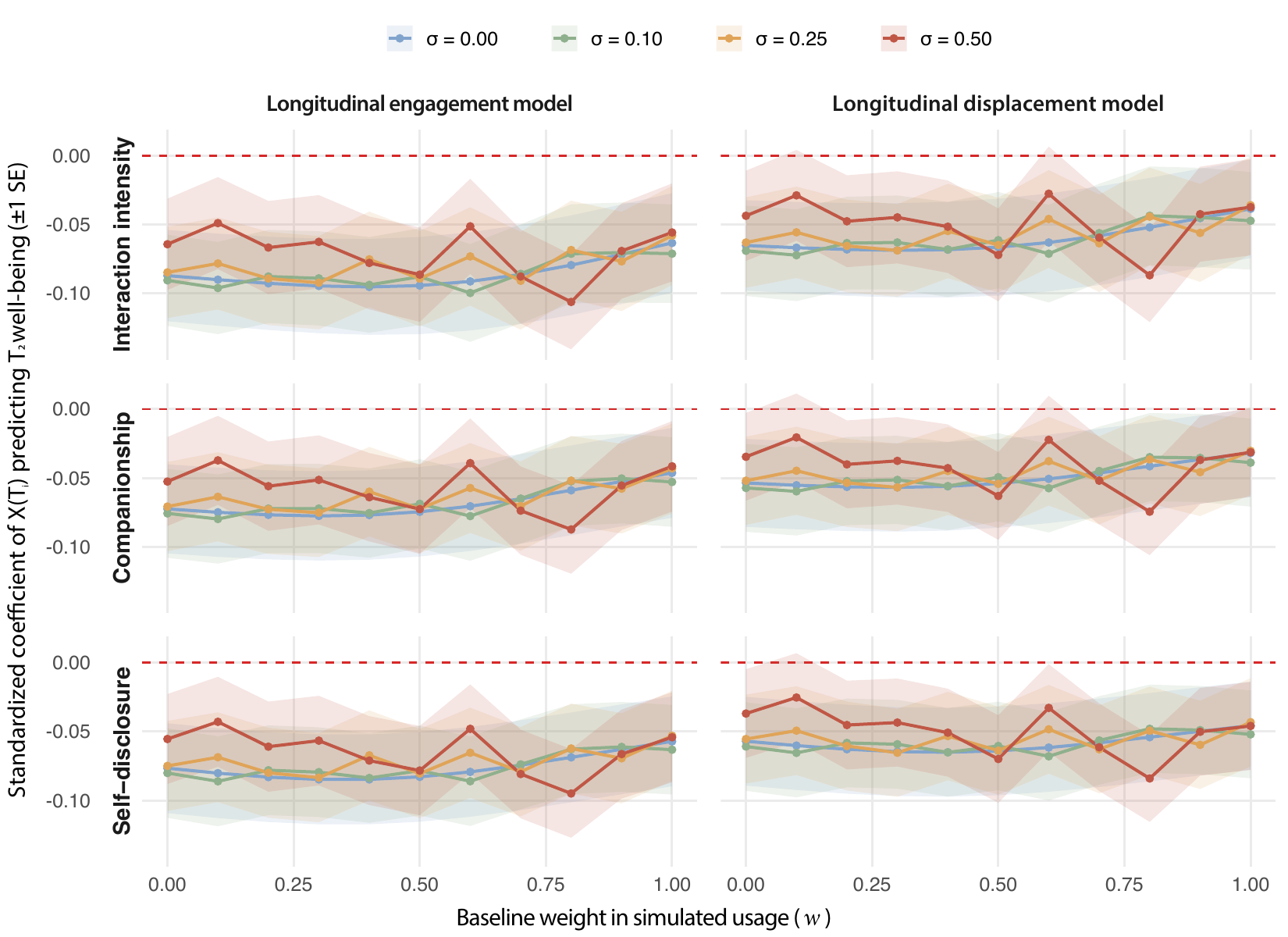}
\caption{\textbf{Sensitivity analysis for the timing of engagement measurement.} Standardized coefficients of the simulated intermediate-time engagement measure $X(T_i)$ predicting follow-up well-being across mixing weights $w$ (Equation~\ref{eq:wavemix}), where $w=0$ corresponds to the follow-up ($T_2$) measure and $w=1$ to the baseline ($T_1$) measure. Shaded bands indicate $\pm 1$ standard error, and $\sigma$ denotes the simulated measurement-noise level.}
\label{fig:sensitivity_sweep}
\end{figure}

%% file: tables/free_text_companionship_predictors_w_discontinued_user.tex
\begin{table*}[ht]
\small
\renewcommand\cellalign{cc}
\setcellgapes{3pt}
\makegapedcells
\renewcommand{\arraystretch}{1}
\resizebox{\textwidth}{!}{%
\begin{tabular}{lcccc}
\toprule
& \makecell[c]{(1) Interaction intensity T$_{2}$} & \makecell[c]{(2) Companionship T$_{2}$} & \makecell[c]{(3) Self-disclosure T$_{2}$} & \makecell[c]{(4) Continued use T$_{2}$} \\
\midrule
\renewcommand{\arraystretch}{3}

Interaction intensity T$_1$
&
\makecell{0.44***\\{}[0.33, 0.55]\\{}$p < 0.001$}
&
\makecell{0.20***\\{}[0.08, 0.32]\\{}$p < 0.001$}
&
\makecell{0.28***\\{}[0.17, 0.40]\\{}$p < 0.001$}
&
\makecell{1.97***\\{}[1.45, 2.66]\\{}$p < 0.001$}
\\

Companionship use T$_1$
&
\makecell{0.03\\{}[-0.06, 0.12]\\{}$p = 0.465$}
&
\makecell{0.22***\\{}[0.13, 0.32]\\{}$p < 0.001$}
&
\makecell{0.04\\{}[-0.05, 0.13]\\{}$p = 0.421$}
&
\makecell{0.98\\{}[0.76, 1.27]\\{}$p = 0.875$}
\\

Self-disclosure T$_1$
&
\makecell{-0.02\\{}[-0.12, 0.08]\\{}$p = 0.664$}
&
\makecell{0.01\\{}[-0.09, 0.11]\\{}$p = 0.825$}
&
\makecell{0.13**\\{}[0.04, 0.23]\\{}$p = 0.008$}
&
\makecell{0.86\\{}[0.66, 1.11]\\{}$p = 0.241$}
\\

Chatbot tenure
&
\makecell{0.01\\{}[-0.07, 0.10]\\{}$p = 0.783$}
&
\makecell{0.02\\{}[-0.07, 0.11]\\{}$p = 0.668$}
&
\makecell{-0.03\\{}[-0.12, 0.06]\\{}$p = 0.562$}
&
\makecell{1.06\\{}[0.83, 1.35]\\{}$p = 0.637$}
\\

Age
&
\makecell{0.11*\\{}[0.00, 0.22]\\{}$p = 0.041$}
&
\makecell{0.16**\\{}[0.04, 0.27]\\{}$p = 0.006$}
&
\makecell{0.08\\{}[-0.03, 0.19]\\{}$p = 0.146$}
&
\makecell{1.18\\{}[0.86, 1.63]\\{}$p = 0.303$}
\\

Male
&
\makecell{0.02\\{}[-0.16, 0.20]\\{}$p = 0.812$}
&
\makecell{0.14\\{}[-0.04, 0.33]\\{}$p = 0.132$}
&
\makecell{0.09\\{}[-0.10, 0.27]\\{}$p = 0.359$}
&
\makecell{1.58\\{}[0.96, 2.61]\\{}$p = 0.073$}
\\

Non-binary
&
\makecell{-0.36*\\{}[-0.67, -0.04]\\{}$p = 0.027$}
&
\makecell{-0.24\\{}[-0.56, 0.09]\\{}$p = 0.156$}
&
\makecell{-0.39*\\{}[-0.71, -0.06]\\{}$p = 0.020$}
&
\makecell{0.35**\\{}[0.16, 0.75]\\{}$p = 0.007$}
\\

Single
&
\makecell{0.09\\{}[-0.09, 0.26]\\{}$p = 0.338$}
&
\makecell{0.08\\{}[-0.10, 0.27]\\{}$p = 0.362$}
&
\makecell{0.07\\{}[-0.10, 0.25]\\{}$p = 0.418$}
&
\makecell{1.01\\{}[0.63, 1.61]\\{}$p = 0.969$}
\\

\midrule
Wald $\chi^2$ & 103.1*** & 73.7*** & 82.6*** & --- \\
Selection correlation ($\rho$) & 0.26 & 0.08 & 0.29 & --- \\
$N$ & 439 & 439 & 439 & 439 \\
\bottomrule
\end{tabular}
}
\vspace{0.5em}

\caption{Longitudinal predictors of interaction intensity, companionship use, self-disclosure, and continued use at follow-up, using participants' free-text relationship descriptions to operationalize companionship. Models (1)-(3) report standardized coefficients predicting each follow-up engagement measure from baseline engagement. Model (4) tests whether these baseline engagement patterns predict whether participants continued using chatbots at follow-up. All estimates are reported with 95\% confidence intervals and two-sided p-values. $^{*}p<0.05$; $^{**}p<0.01$; $^{***}p<0.001$.}
\label{tab:free_text_predictors_wdu}
\end{table*}

%% file: tables/sem_persistence_combined_w_discontinued_user.tex
\begin{table*}[!htbp]
\small
\renewcommand\cellalign{cc}
\renewcommand{\arraystretch}{1}
\setcellgapes{2pt}
\makegapedcells

\resizebox{\textwidth}{!}{%
\begin{tabular}{@{}lcccccccc@{}}
\toprule
 & \multicolumn{2}{c}{\makecell{Interaction intensity}}
 & \multicolumn{2}{c}{\makecell{Companionship (forced choice)}}
 & \multicolumn{2}{c}{\makecell{Companionship (free-text)}}
 & \multicolumn{2}{c}{\makecell{Self-disclosure}} \\
\cmidrule(lr){2-3}\cmidrule(lr){4-5}\cmidrule(lr){6-7}\cmidrule(lr){8-9}
 & \makecell[c]{(1) \\ Interaction intensity \\ T$_{2}$}
& \makecell[c]{(2) \\ Well-being \\ T$_{2}$}
& \makecell[c]{(3) \\ Companionship \\ T$_{2}$}
& \makecell[c]{(4) \\ Well-being \\ T$_{2}$}
& \makecell[c]{(5) \\ Companionship \\ T$_{2}$}
& \makecell[c]{(6) \\ Well-being \\ T$_{2}$}
& \makecell[c]{(7) \\ Self-disclosure \\ T$_{2}$}
& \makecell[c]{(8) \\ Well-being \\ T$_{2}$} \\
\midrule
\renewcommand{\arraystretch}{3}

\makecell[l]{Interaction \\ intensity T$_1$} 
& \makecell{0.42***\\{}[0.32, 0.51]\\{}$p < 0.001$}
& \makecell{-0.03\\{}[-0.13, 0.06]\\{}$p = 0.509$}
& ---
& ---
& ---
& ---
& ---
& --- \\

\makecell[l]{Interaction \\ intensity T$_2$}
& ---
& \makecell{-0.08*\\{}[-0.14, -0.01]\\{}$p = 0.025$}
& ---
& ---
& ---
& ---
& ---
& --- \\

Companionship T$_{1}$
& ---
& ---
& \makecell{0.23***\\{}[0.13, 0.33]\\{}$p < 0.001$}
& \makecell{-0.03\\{}[-0.09, 0.03]\\{}$p = 0.274$}
& \makecell{0.28***\\{}[0.19, 0.37]\\{}$p < 0.001$}
& \makecell{-0.03\\{}[-0.09, 0.03]\\{}$p = 0.367$}
& ---
& --- \\

Companionship T$_{2}$
& ---
& ---
& ---
& \makecell{-0.06\\{}[-0.13, 0.01]\\{}$p = 0.101$}
& ---
& \makecell{-0.06*\\{}[-0.13, -0.00]\\{}$p = 0.040$}
& ---
& --- \\

Self-disclosure T$_{1}$
& ---
& ---
& ---
& ---
& ---
& ---
& \makecell{0.29***\\{}[0.21, 0.38]\\{}$p < 0.001$}
& \makecell{-0.04\\{}[-0.12, 0.05]\\{}$p = 0.369$} \\

Self-disclosure T$_{2}$
& ---
& ---
& ---
& ---
& ---
& ---
& ---
& \makecell{-0.07\\{}[-0.13, 0.00]\\{}$p = 0.056$} \\

Well-being T$_{1}$ & --- &
\makecell{0.75***\\{}[0.68, 0.81]\\{}$p < 0.001$} &
--- &
\makecell{0.76***\\{}[0.70, 0.82]\\{}$p < 0.001$} &
--- &
\makecell{0.75***\\{}[0.69, 0.82]\\{}$p < 0.001$} &
--- &
\makecell{0.75***\\{}[0.69, 0.81]\\{}$p < 0.001$} \\

Chatbot tenure & \makecell{0.01\\{}[-0.07, 0.10]\\{}$p = 0.758$} &
\makecell{0.02\\{}[-0.04, 0.08]\\{}$p = 0.534$} &
\makecell{0.08\\{}[-0.01, 0.17]\\{}$p = 0.069$} &
\makecell{0.01\\{}[-0.05, 0.07]\\{}$p = 0.706$} &
\makecell{0.06\\{}[-0.03, 0.15]\\{}$p = 0.181$} &
\makecell{0.01\\{}[-0.05, 0.07]\\{}$p = 0.661$} &
\makecell{0.02\\{}[-0.07, 0.11]\\{}$p = 0.636$} &
\makecell{0.01\\{}[-0.05, 0.07]\\{}$p = 0.665$} \\

Age & \makecell{0.09\\{}[-0.01, 0.18]\\{}$p = 0.065$} &
\makecell{0.02\\{}[-0.05, 0.09]\\{}$p = 0.537$} &
\makecell{0.16**\\{}[0.06, 0.25]\\{}$p = 0.001$} &
\makecell{0.01\\{}[-0.05, 0.08]\\{}$p = 0.704$} &
\makecell{0.16***\\{}[0.07, 0.26]\\{}$p < 0.001$} &
\makecell{0.01\\{}[-0.05, 0.08]\\{}$p = 0.757$} &
\makecell{0.09\\{}[-0.01, 0.19]\\{}$p = 0.068$} &
\makecell{0.02\\{}[-0.05, 0.08]\\{}$p = 0.645$} \\

Male & \makecell{0.01\\{}[-0.07, 0.10]\\{}$p = 0.750$} &
\makecell{-0.01\\{}[-0.08, 0.06]\\{}$p = 0.846$} &
\makecell{0.04\\{}[-0.05, 0.13]\\{}$p = 0.341$} &
\makecell{0.00\\{}[-0.06, 0.07]\\{}$p = 0.935$} &
\makecell{0.04\\{}[-0.05, 0.13]\\{}$p = 0.390$} &
\makecell{0.00\\{}[-0.06, 0.07]\\{}$p = 0.911$} &
\makecell{0.02\\{}[-0.06, 0.11]\\{}$p = 0.578$} &
\makecell{-0.01\\{}[-0.07, 0.06]\\{}$p = 0.874$} \\

Non-binary & \makecell{-0.10*\\{}[-0.20, -0.01]\\{}$p = 0.031$} &
\makecell{-0.03\\{}[-0.09, 0.03]\\{}$p = 0.278$} &
\makecell{-0.06\\{}[-0.16, 0.03]\\{}$p = 0.187$} &
\makecell{-0.03\\{}[-0.09, 0.03]\\{}$p = 0.317$} &
\makecell{-0.05\\{}[-0.14, 0.04]\\{}$p = 0.266$} &
\makecell{-0.04\\{}[-0.09, 0.02]\\{}$p = 0.233$} &
\makecell{-0.09\\{}[-0.19, 0.01]\\{}$p = 0.071$} &
\makecell{-0.04\\{}[-0.09, 0.02]\\{}$p = 0.222$} \\

Single & \makecell{0.04\\{}[-0.04, 0.13]\\{}$p = 0.334$} &
\makecell{-0.00\\{}[-0.06, 0.06]\\{}$p = 0.959$} &
\makecell{0.04\\{}[-0.05, 0.13]\\{}$p = 0.416$} &
\makecell{-0.01\\{}[-0.08, 0.05]\\{}$p = 0.689$} &
\makecell{0.06\\{}[-0.03, 0.15]\\{}$p = 0.210$} &
\makecell{-0.01\\{}[-0.07, 0.06]\\{}$p = 0.845$} &
\makecell{0.06\\{}[-0.03, 0.15]\\{}$p = 0.199$} &
\makecell{-0.01\\{}[-0.07, 0.06]\\{}$p = 0.863$} \\

Inverse Mills ratio & \makecell{0.08\\{}[-0.02, 0.18]\\{}$p = 0.102$} &
\makecell{0.01\\{}[-0.07, 0.09]\\{}$p = 0.810$} &
\makecell{0.21***\\{}[0.12, 0.31]\\{}$p < 0.001$} &
\makecell{-0.01\\{}[-0.09, 0.06]\\{}$p = 0.757$} &
\makecell{0.09\\{}[-0.01, 0.19]\\{}$p = 0.080$} &
\makecell{-0.01\\{}[-0.09, 0.06]\\{}$p = 0.738$} &
\makecell{0.18***\\{}[0.09, 0.27]\\{}$p < 0.001$} &
\makecell{-0.01\\{}[-0.08, 0.06]\\{}$p = 0.816$} \\

\midrule

Direct effect & --- &
\makecell{-0.03\\{}[-0.13, 0.06]\\{}$p = 0.509$} &
--- &
\makecell{-0.03\\{}[-0.09, 0.03]\\{}$p = 0.274$} &
--- &
\makecell{-0.03\\{}[-0.09, 0.03]\\{}$p = 0.367$} &
--- &
\makecell{-0.04\\{}[-0.12, 0.05]\\{}$p = 0.369$} \\

\makecell[l]{Indirect effect \\ via $T_2$ interaction \\ pattern} & --- &
\makecell{-0.03*\\{}[-0.07, -0.00]\\{}$p = 0.025$} &
--- &
\makecell{-0.01\\{}[-0.03, 0.00]\\{}$p = 0.103$} &
--- &
\makecell{-0.02*\\{}[-0.04, -0.00]\\{}$p = 0.037$} &
--- &
\makecell{-0.02*\\{}[-0.04, -0.00]\\{}$p = 0.048$} \\

\midrule
$R^2$ & 0.235 & 0.579 & 0.141 & 0.575 & 0.128 & 0.577 & 0.152 & 0.578 \\
$N$ & 439 & 439 & 439 & 439 & 439 & 439 & 439 & 439 \\
\bottomrule
\end{tabular}
}
\vspace{0.5em}
\caption{\textbf{Longitudinal associations between sustained social chatbot engagement and subsequent well-being.} 
Models (1)-(2) examine sustained interaction intensity, Models (3)-(6) examine sustained companionship use using the self-report and free-text operationalizations, and Models (7)-(8) examine sustained self-disclosure. For each engagement measure, the first model predicts follow-up engagement from baseline engagement, and the second predicts follow-up well-being from baseline and follow-up engagement. Indirect effects capture the association between baseline engagement and follow-up well-being through follow-up engagement and are estimated using percentile bootstrap confidence intervals (5,000 resamples). All estimates are reported as standardized coefficients with 95\% confidence intervals and two-sided $p$-values. $^{*}p<0.05$; $^{**}p<0.01$; $^{***}p<0.001$.
}
\label{tab:sem_persistence_combined_wdu}
\end{table*}

%% file: tables/sem_displacement_combined_w_discontinued_user.tex
\begin{table*}[h]
\small
\renewcommand\cellalign{cc}
\renewcommand{\arraystretch}{1}
\setcellgapes{2pt}
\makegapedcells

\resizebox{\textwidth}{!}{%
\begin{tabular}{@{}lccccccccc@{}}
\toprule
 & \multicolumn{3}{c}{\makecell{Interaction intensity}}
 & \multicolumn{3}{c}{\makecell{Companionship (free-text)}}
 & \multicolumn{3}{c}{\makecell{Self-disclosure}} \\
\cmidrule(lr){2-4}\cmidrule(lr){5-7}\cmidrule(lr){8-10}
 & \makecell[c]{(1) \\ Interaction \\ intensity \\ T$_2$}
 & \makecell[c]{(2) \\ Time in \\ person \\ T$_{2}$} 
 & \makecell[c]{(3) \\ Well-being \\ T$_{2}$}
 & \makecell[c]{(4) \\ Companionship \\ T$_{2}$} 
 & \makecell[c]{(5) \\ Time in \\ person \\ T$_{2}$} 
 & \makecell[c]{(6) \\ Well-being \\ T$_{2}$}
 & \makecell[c]{(7) \\ Self-disclosure \\ T$_{2}$} 
 & \makecell[c]{(8) \\ Time in \\ person \\ T$_{2}$} 
 & \makecell[c]{(9) \\ Well-being \\ T$_{2}$} \\
\midrule
\renewcommand{\arraystretch}{3}

\makecell[l]{Interaction \\ intensity T$_1$}
& \makecell{0.42***\\{}[0.32, 0.51]\\{}$p < 0.001$}
& ---
& \makecell{-0.01\\{}[-0.11, 0.08]\\{}$p = 0.759$}
& ---
& ---
& ---
& ---
& ---
& --- \\

\makecell[l]{Interaction \\ intensity T$_2$}
& ---
& \makecell{-0.17***\\{}[-0.27, -0.08]\\{}$p < 0.001$}
& \makecell{-0.06\\{}[-0.13, 0.01]\\{}$p = 0.074$}
& ---
& ---
& ---
& ---
& ---
& --- \\

\makecell[l]{Companionship \\ T$_{1}$}
& ---
& ---
& ---
& \makecell{0.28***\\{}[0.19, 0.37]\\{}$p < 0.001$}
& ---
& \makecell{-0.02\\{}[-0.08, 0.04]\\{}$p = 0.539$}
& ---
& ---
& --- \\

\makecell[l]{Companionship \\ T$_{2}$}
& ---
& ---
& ---
& ---
& \makecell{-0.14**\\{}[-0.23, -0.05]\\{}$p = 0.002$}
& \makecell{-0.05\\{}[-0.11, 0.01]\\{}$p = 0.118$}
& ---
& ---
& --- \\

Self-disclosure T$_{1}$
& ---
& ---
& ---
& ---
& ---
& ---
& \makecell{0.29***\\{}[0.21, 0.38]\\{}$p < 0.001$}
& ---
& \makecell{-0.03\\{}[-0.12, 0.05]\\{}$p = 0.442$} \\

Self-disclosure T$_{2}$
& ---
& ---
& ---
& ---
& ---
& ---
& ---
& \makecell{-0.15**\\{}[-0.24, -0.05]\\{}$p = 0.002$}
& \makecell{-0.05\\{}[-0.12, 0.02]\\{}$p = 0.152$} \\

Time in person T$_{2}$
& ---
& ---
& \makecell{0.14***\\{}[0.08, 0.21]\\{}$p < 0.001$}
& ---
& ---
& \makecell{0.15***\\{}[0.08, 0.22]\\{}$p < 0.001$}
& ---
& ---
& \makecell{0.15***\\{}[0.08, 0.21]\\{}$p < 0.001$} \\

Well-being T$_{1}$
& ---
& ---
& \makecell{0.73***\\{}[0.67, 0.80]\\{}$p < 0.001$}
& ---
& ---
& \makecell{0.74***\\{}[0.67, 0.80]\\{}$p < 0.001$}
& ---
& ---
& \makecell{0.74***\\{}[0.68, 0.80]\\{}$p < 0.001$} \\

Chatbot tenure
& \makecell{0.01\\{}[-0.07, 0.10]\\{}$p = 0.757$}
& \makecell{0.03\\{}[-0.06, 0.12]\\{}$p = 0.527$}
& \makecell{0.01\\{}[-0.05, 0.07]\\{}$p = 0.684$}
& \makecell{0.06\\{}[-0.03, 0.15]\\{}$p = 0.181$}
& \makecell{0.02\\{}[-0.07, 0.11]\\{}$p = 0.617$}
& \makecell{0.01\\{}[-0.05, 0.07]\\{}$p = 0.756$}
& \makecell{0.02\\{}[-0.07, 0.11]\\{}$p = 0.635$}
& \makecell{0.02\\{}[-0.07, 0.11]\\{}$p = 0.686$}
& \makecell{0.01\\{}[-0.05, 0.07]\\{}$p = 0.742$} \\

Age
& \makecell{0.09\\{}[-0.01, 0.18]\\{}$p = 0.065$}
& \makecell{-0.06\\{}[-0.15, 0.03]\\{}$p = 0.178$}
& \makecell{0.03\\{}[-0.03, 0.10]\\{}$p = 0.347$}
& \makecell{0.16***\\{}[0.07, 0.26]\\{}$p < 0.001$}
& \makecell{-0.07\\{}[-0.16, 0.02]\\{}$p = 0.137$}
& \makecell{0.03\\{}[-0.04, 0.09]\\{}$p = 0.444$}
& \makecell{0.09\\{}[-0.01, 0.19]\\{}$p = 0.068$}
& \makecell{-0.07\\{}[-0.16, 0.02]\\{}$p = 0.122$}
& \makecell{0.03\\{}[-0.04, 0.10]\\{}$p = 0.378$} \\

Male
& \makecell{0.01\\{}[-0.07, 0.10]\\{}$p = 0.750$}
& \makecell{-0.22***\\{}[-0.31, -0.13]\\{}$p < 0.001$}
& \makecell{0.03\\{}[-0.04, 0.10]\\{}$p = 0.467$}
& \makecell{0.04\\{}[-0.05, 0.13]\\{}$p = 0.390$}
& \makecell{-0.21***\\{}[-0.30, -0.11]\\{}$p < 0.001$}
& \makecell{0.03\\{}[-0.04, 0.10]\\{}$p = 0.335$}
& \makecell{0.02\\{}[-0.06, 0.11]\\{}$p = 0.577$}
& \makecell{-0.21***\\{}[-0.30, -0.12]\\{}$p < 0.001$}
& \makecell{0.03\\{}[-0.05, 0.10]\\{}$p = 0.475$} \\

Non-binary
& \makecell{-0.10*\\{}[-0.20, -0.01]\\{}$p = 0.031$}
& \makecell{-0.00\\{}[-0.11, 0.11]\\{}$p = 0.976$}
& \makecell{-0.03\\{}[-0.09, 0.03]\\{}$p = 0.274$}
& \makecell{-0.05\\{}[-0.14, 0.04]\\{}$p = 0.266$}
& \makecell{0.00\\{}[-0.11, 0.11]\\{}$p = 0.980$}
& \makecell{-0.03\\{}[-0.09, 0.02]\\{}$p = 0.259$}
& \makecell{-0.09\\{}[-0.19, 0.01]\\{}$p = 0.070$}
& \makecell{-0.00\\{}[-0.11, 0.10]\\{}$p = 0.963$}
& \makecell{-0.03\\{}[-0.09, 0.02]\\{}$p = 0.248$} \\

Single
& \makecell{0.04\\{}[-0.04, 0.13]\\{}$p = 0.335$}
& \makecell{-0.04\\{}[-0.13, 0.05]\\{}$p = 0.353$}
& \makecell{0.00\\{}[-0.06, 0.06]\\{}$p = 0.902$}
& \makecell{0.06\\{}[-0.03, 0.15]\\{}$p = 0.209$}
& \makecell{-0.05\\{}[-0.14, 0.04]\\{}$p = 0.305$}
& \makecell{0.00\\{}[-0.06, 0.06]\\{}$p = 0.972$}
& \makecell{0.06\\{}[-0.03, 0.15]\\{}$p = 0.198$}
& \makecell{-0.05\\{}[-0.14, 0.04]\\{}$p = 0.311$}
& \makecell{0.00\\{}[-0.06, 0.06]\\{}$p = 0.960$} \\

Inverse Mills ratio
& \makecell{0.08\\{}[-0.02, 0.18]\\{}$p = 0.102$}
& \makecell{0.18***\\{}[0.08, 0.27]\\{}$p < 0.001$}
& \makecell{-0.01\\{}[-0.09, 0.07]\\{}$p = 0.859$}
& \makecell{0.09\\{}[-0.01, 0.19]\\{}$p = 0.081$}
& \makecell{0.15**\\{}[0.06, 0.25]\\{}$p = 0.001$}
& \makecell{-0.02\\{}[-0.09, 0.05]\\{}$p = 0.563$}
& \makecell{0.18***\\{}[0.09, 0.27]\\{}$p < 0.001$}
& \makecell{0.17***\\{}[0.07, 0.26]\\{}$p < 0.001$}
& \makecell{-0.02\\{}[-0.09, 0.05]\\{}$p = 0.624$} \\

\midrule

Direct effect
& ---
& ---
& \makecell{-0.06\\{}[-0.13, 0.01]\\{}$p = 0.074$}
& ---
& ---
& \makecell{-0.05\\{}[-0.11, 0.01]\\{}$p = 0.118$}
& ---
& ---
& \makecell{-0.05\\{}[-0.12, 0.02]\\{}$p = 0.152$} \\

\makecell[l]{Indirect effect via \\ $T_2$ human social \\ interaction}
& ---
& ---
& \makecell{-0.02**\\{}[-0.05, -0.01]\\{}$p = 0.001$}
& ---
& ---
& \makecell{-0.02**\\{}[-0.04, -0.01]\\{}$p = 0.004$}
& ---
& ---
& \makecell{-0.02**\\{}[-0.04, -0.01]\\{}$p = 0.002$} \\

\midrule
$R^2$ & 0.235 & 0.096 & 0.584 & 0.128 & 0.089 & 0.582 & 0.152 & 0.089 & 0.583 \\
$N$ & 439 & 439 & 439 & 439 & 439 & 439 & 439 & 439 & 439 \\
\bottomrule
\end{tabular}
}

\vspace{0.5em}
\caption{\textbf{Longitudinal associations between social chatbot engagement, human social interaction, and subsequent well-being.} Models (1)-(3) examine interaction intensity, Models (4)-(6) companionship use measured from free-text responses, and Models (7)-(9) self-disclosure. 
For each engagement measure, the models predict follow-up engagement, in-person social interaction, and well-being, respectively. 
Indirect effects through in-person social interaction are estimated using percentile bootstrap confidence intervals (5,000 resamples). All estimates are standardized coefficients with 95\% confidence intervals and two-sided $p$-values. $^{*}p<0.05$; $^{**}p<0.01$; $^{***}p<0.001$.}
\label{tab:sem_displacement_combined_wdu}
\end{table*}

%% file: tables/cross_pred_companionship_predictors_331.tex
\begin{table*}[ht]
\small
\renewcommand\cellalign{cc}
\setcellgapes{3pt}
\makegapedcells
\renewcommand{\arraystretch}{1}
\resizebox{\textwidth}{!}{%
\begin{tabular}{@{}lcccccc@{}}
\toprule
 & \multicolumn{3}{c}{\makecell{Forced-choice companionship}} & \multicolumn{3}{c}{\makecell{Free-text companionship}} \\
\cmidrule(lr){2-4}\cmidrule(lr){5-7}
 & \makecell[c]{(1) \\ Interaction intensity \\ T$_{2}$} & \makecell[c]{(2) \\ Companionship \\ T$_{2}$} & \makecell[c]{(3) \\ Self-disclosure \\ T$_{2}$} & \makecell[c]{(4) \\ Interaction intensity \\ T$_{2}$} & \makecell[c]{(5) \\ Companionship \\ T$_{2}$} & \makecell[c]{(6) \\ Self-disclosure \\ T$_{2}$} \\
\midrule
\renewcommand{\arraystretch}{3}

Interaction intensity T$_{1}$ & \makecell{0.52***\\{}[0.40, 0.64]\\{}$p < 0.001$} &
\makecell{0.28***\\{}[0.15, 0.40]\\{}$p < 0.001$} &
\makecell{0.19**\\{}[0.07, 0.31]\\{}$p = 0.003$} &
\makecell{0.54***\\{}[0.42, 0.66]\\{}$p < 0.001$} &
\makecell{0.10\\{}[-0.03, 0.24]\\{}$p = 0.141$} &
\makecell{0.19**\\{}[0.07, 0.31]\\{}$p = 0.003$} \\

Companionship T$_{1}$ & \makecell{0.55***\\{}[0.27, 0.84]\\{}$p < 0.001$} &
\makecell{0.63***\\{}[0.32, 0.93]\\{}$p < 0.001$} &
\makecell{0.20\\{}[-0.10, 0.49]\\{}$p = 0.187$} &
\makecell{0.08\\{}[-0.02, 0.18]\\{}$p = 0.125$} &
\makecell{0.32***\\{}[0.21, 0.43]\\{}$p < 0.001$} &
\makecell{0.08\\{}[-0.02, 0.18]\\{}$p = 0.126$} \\

Self-disclosure T$_{1}$ & \makecell{-0.04\\{}[-0.14, 0.06]\\{}$p = 0.470$} &
\makecell{0.17**\\{}[0.06, 0.27]\\{}$p = 0.003$} &
\makecell{0.37***\\{}[0.26, 0.47]\\{}$p < 0.001$} &
\makecell{-0.04\\{}[-0.14, 0.07]\\{}$p = 0.504$} &
\makecell{0.04\\{}[-0.07, 0.16]\\{}$p = 0.455$} &
\makecell{0.35***\\{}[0.24, 0.46]\\{}$p < 0.001$} \\

Chatbot tenure & \makecell{0.00\\{}[-0.09, 0.10]\\{}$p = 0.933$} &
\makecell{0.01\\{}[-0.09, 0.11]\\{}$p = 0.831$} &
\makecell{-0.06\\{}[-0.16, 0.04]\\{}$p = 0.230$} &
\makecell{0.00\\{}[-0.09, 0.10]\\{}$p = 0.939$} &
\makecell{0.04\\{}[-0.07, 0.14]\\{}$p = 0.520$} &
\makecell{-0.06\\{}[-0.16, 0.04]\\{}$p = 0.223$} \\

Age & \makecell{0.16**\\{}[0.05, 0.27]\\{}$p = 0.005$} &
\makecell{0.15*\\{}[0.03, 0.27]\\{}$p = 0.016$} &
\makecell{0.08\\{}[-0.04, 0.20]\\{}$p = 0.177$} &
\makecell{0.17**\\{}[0.06, 0.29]\\{}$p = 0.003$} &
\makecell{0.20**\\{}[0.07, 0.33]\\{}$p = 0.002$} &
\makecell{0.10\\{}[-0.02, 0.22]\\{}$p = 0.112$} \\

Male & \makecell{-0.22*\\{}[-0.40, -0.04]\\{}$p = 0.020$} &
\makecell{0.17\\{}[-0.03, 0.36]\\{}$p = 0.096$} &
\makecell{-0.07\\{}[-0.26, 0.12]\\{}$p = 0.479$} &
\makecell{-0.24*\\{}[-0.42, -0.05]\\{}$p = 0.014$} &
\makecell{0.07\\{}[-0.13, 0.28]\\{}$p = 0.484$} &
\makecell{-0.08\\{}[-0.27, 0.11]\\{}$p = 0.406$} \\

Non-binary & \makecell{-0.31\\{}[-0.70, 0.08]\\{}$p = 0.114$} &
\makecell{-0.24\\{}[-0.65, 0.17]\\{}$p = 0.254$} &
\makecell{-0.28\\{}[-0.68, 0.12]\\{}$p = 0.168$} &
\makecell{-0.18\\{}[-0.57, 0.21]\\{}$p = 0.364$} &
\makecell{-0.06\\{}[-0.49, 0.38]\\{}$p = 0.803$} &
\makecell{-0.22\\{}[-0.61, 0.18]\\{}$p = 0.288$} \\

Single & \makecell{0.18\\{}[-0.01, 0.37]\\{}$p = 0.057$} &
\makecell{-0.03\\{}[-0.23, 0.17]\\{}$p = 0.758$} &
\makecell{0.08\\{}[-0.11, 0.27]\\{}$p = 0.418$} &
\makecell{0.15\\{}[-0.04, 0.34]\\{}$p = 0.122$} &
\makecell{0.10\\{}[-0.11, 0.31]\\{}$p = 0.351$} &
\makecell{0.07\\{}[-0.12, 0.26]\\{}$p = 0.476$} \\

\midrule
\renewcommand{\arraystretch}{1}
Wald $\chi^2$ & 170.7*** & 111.7*** & 137.3*** & 153.8*** & 67.5*** & 138.5*** \\
Selection correlation ($\rho$) & 0.31 & 0.32 & 0.23 & 0.24 & 0.00 & 0.22 \\
$N$ & 331 & 331 & 331 & 331 & 331 & 331 \\
\bottomrule
\end{tabular}
}
\vspace{0.5em}

\caption{Longitudinal predictors of subsequent chatbot engagement among participants who continued using chatbots at follow-up ($N=331$), using both participants' force-choice selection and free-text relationship descriptions to operationalize companionship use. Results report standardized coefficients predicting each follow-up engagement measure from baseline engagement. All estimates are reported with 95\% confidence intervals and two-sided $p$-values.$^{*}p<0.05$; $^{**}p<0.01$; $^{***}p<0.001$.}
\label{tab:cross_pred_predictors_331_combined}
\end{table*}

%% file: tables/sem_persistence_combined_331.tex
\begin{table*}[ht]
\small
\renewcommand\cellalign{cc}
\renewcommand{\arraystretch}{1}
\setcellgapes{2pt}
\makegapedcells

\resizebox{\textwidth}{!}{%
\begin{tabular}{@{}lcccccccc@{}}
\toprule
 & \multicolumn{2}{c}{\makecell{Interaction intensity}}
 & \multicolumn{2}{c}{\makecell{Companionship (forced choice)}}
 & \multicolumn{2}{c}{\makecell{Companionship (free-text)}}
 & \multicolumn{2}{c}{\makecell{Self-disclosure}} \\
\cmidrule(lr){2-3}\cmidrule(lr){4-5}\cmidrule(lr){6-7}\cmidrule(lr){8-9}
 & \makecell[c]{(1) \\ Interaction intensity \\ T$_{2}$}
& \makecell[c]{(2) \\ Well-being \\ T$_{2}$}
& \makecell[c]{(3) \\ Companionship \\ T$_{2}$}
& \makecell[c]{(4) \\ Well-being \\ T$_{2}$}
& \makecell[c]{(5) \\ Companionship \\ T$_{2}$}
& \makecell[c]{(6) \\ Well-being \\ T$_{2}$}
& \makecell[c]{(7) \\ Self-disclosure \\ T$_{2}$}
& \makecell[c]{(8) \\ Well-being \\ T$_{2}$} \\
\midrule
\renewcommand{\arraystretch}{3}

\makecell[l]{Interaction \\ intensity T$_{1}$} & \makecell{0.51***\\{}[0.42, 0.60]\\{}$p < 0.001$} &
\makecell{-0.02\\{}[-0.13, 0.09]\\{}$p = 0.757$} &
--- &
--- &
--- &
--- &
--- &
--- \\

Companionship T$_{1}$ & --- &
--- &
\makecell{0.31***\\{}[0.21, 0.41]\\{}$p < 0.001$} &
\makecell{-0.05\\{}[-0.12, 0.03]\\{}$p = 0.213$} &
\makecell{0.37***\\{}[0.27, 0.47]\\{}$p < 0.001$} &
\makecell{-0.01\\{}[-0.08, 0.07]\\{}$p = 0.874$} &
--- &
--- \\

Self-disclosure T$_{1}$ & --- &
--- &
--- &
--- &
--- &
--- &
\makecell{0.50***\\{}[0.40, 0.59]\\{}$p < 0.001$} &
\makecell{0.01\\{}[-0.10, 0.11]\\{}$p = 0.899$} \\

Focal predictor T$_{2}$ & --- &
\makecell{-0.12**\\{}[-0.21, -0.04]\\{}$p = 0.003$} &
--- &
\makecell{-0.07\\{}[-0.15, 0.02]\\{}$p = 0.109$} &
--- &
\makecell{-0.08*\\{}[-0.16, -0.01]\\{}$p = 0.031$} &
--- &
\makecell{-0.12*\\{}[-0.22, -0.02]\\{}$p = 0.020$} \\

Well-being T$_{1}$ & --- &
\makecell{0.74***\\{}[0.66, 0.81]\\{}$p < 0.001$} &
--- &
\makecell{0.75***\\{}[0.67, 0.82]\\{}$p < 0.001$} &
--- &
\makecell{0.75***\\{}[0.67, 0.82]\\{}$p < 0.001$} &
--- &
\makecell{0.75***\\{}[0.68, 0.82]\\{}$p < 0.001$} \\

Chatbot tenure & \makecell{0.00\\{}[-0.09, 0.10]\\{}$p = 0.928$} &
\makecell{0.04\\{}[-0.04, 0.11]\\{}$p = 0.321$} &
\makecell{0.07\\{}[-0.03, 0.18]\\{}$p = 0.167$} &
\makecell{0.03\\{}[-0.05, 0.10]\\{}$p = 0.467$} &
\makecell{0.06\\{}[-0.05, 0.16]\\{}$p = 0.283$} &
\makecell{0.03\\{}[-0.05, 0.10]\\{}$p = 0.493$} &
\makecell{-0.02\\{}[-0.12, 0.07]\\{}$p = 0.639$} &
\makecell{0.02\\{}[-0.05, 0.09]\\{}$p = 0.588$} \\

Age & \makecell{0.13**\\{}[0.05, 0.21]\\{}$p = 0.002$} &
\makecell{0.04\\{}[-0.04, 0.11]\\{}$p = 0.367$} &
\makecell{0.17***\\{}[0.08, 0.27]\\{}$p < 0.001$} &
\makecell{0.02\\{}[-0.05, 0.09]\\{}$p = 0.604$} &
\makecell{0.19***\\{}[0.09, 0.28]\\{}$p < 0.001$} &
\makecell{0.02\\{}[-0.06, 0.09]\\{}$p = 0.684$} &
\makecell{0.08*\\{}[0.00, 0.16]\\{}$p = 0.046$} &
\makecell{0.02\\{}[-0.06, 0.10]\\{}$p = 0.610$} \\

Male & \makecell{-0.11*\\{}[-0.20, -0.02]\\{}$p = 0.015$} &
\makecell{-0.01\\{}[-0.09, 0.07]\\{}$p = 0.815$} &
\makecell{0.03\\{}[-0.07, 0.13]\\{}$p = 0.559$} &
\makecell{0.01\\{}[-0.06, 0.09]\\{}$p = 0.756$} &
\makecell{0.02\\{}[-0.09, 0.12]\\{}$p = 0.732$} &
\makecell{0.02\\{}[-0.06, 0.09]\\{}$p = 0.700$} &
\makecell{-0.04\\{}[-0.14, 0.05]\\{}$p = 0.380$} &
\makecell{0.00\\{}[-0.08, 0.08]\\{}$p = 0.989$} \\

Non-binary & \makecell{-0.05\\{}[-0.16, 0.06]\\{}$p = 0.362$} &
\makecell{-0.00\\{}[-0.06, 0.05]\\{}$p = 0.886$} &
\makecell{-0.05\\{}[-0.17, 0.07]\\{}$p = 0.423$} &
\makecell{0.00\\{}[-0.05, 0.05]\\{}$p = 0.900$} &
\makecell{-0.00\\{}[-0.11, 0.10]\\{}$p = 0.967$} &
\makecell{-0.01\\{}[-0.06, 0.05]\\{}$p = 0.831$} &
\makecell{-0.05\\{}[-0.15, 0.05]\\{}$p = 0.335$} &
\makecell{-0.01\\{}[-0.06, 0.05]\\{}$p = 0.768$} \\

Single & \makecell{0.07\\{}[-0.02, 0.16]\\{}$p = 0.142$} &
\makecell{-0.02\\{}[-0.10, 0.05]\\{}$p = 0.548$} &
\makecell{-0.00\\{}[-0.11, 0.11]\\{}$p = 0.963$} &
\makecell{-0.04\\{}[-0.12, 0.04]\\{}$p = 0.304$} &
\makecell{0.05\\{}[-0.05, 0.16]\\{}$p = 0.337$} &
\makecell{-0.03\\{}[-0.11, 0.05]\\{}$p = 0.454$} &
\makecell{0.03\\{}[-0.07, 0.13]\\{}$p = 0.530$} &
\makecell{-0.03\\{}[-0.10, 0.05]\\{}$p = 0.483$} \\

Inverse Mills ratio & \makecell{0.04\\{}[-0.04, 0.13]\\{}$p = 0.329$} &
\makecell{-0.00\\{}[-0.08, 0.08]\\{}$p = 0.941$} &
\makecell{0.11\\{}[-0.00, 0.22]\\{}$p = 0.059$} &
\makecell{-0.03\\{}[-0.11, 0.06]\\{}$p = 0.546$} &
\makecell{0.01\\{}[-0.09, 0.12]\\{}$p = 0.795$} &
\makecell{-0.03\\{}[-0.10, 0.05]\\{}$p = 0.532$} &
\makecell{0.08\\{}[-0.02, 0.18]\\{}$p = 0.111$} &
\makecell{-0.02\\{}[-0.09, 0.06]\\{}$p = 0.686$} \\

\midrule

Direct effect & --- &
\makecell{-0.02\\{}[-0.13, 0.09]\\{}$p = 0.757$} &
--- &
\makecell{-0.05\\{}[-0.12, 0.03]\\{}$p = 0.213$} &
--- &
\makecell{-0.01\\{}[-0.08, 0.07]\\{}$p = 0.874$} &
--- &
\makecell{0.01\\{}[-0.10, 0.11]\\{}$p = 0.899$} \\

\makecell[l]{Indirect effect \\ via $T_2$ interaction \\ pattern} & --- &
\makecell{-0.06**\\{}[-0.12, -0.02]\\{}$p = 0.001$} &
--- &
\makecell{-0.02\\{}[-0.05, 0.01]\\{}$p = 0.117$} &
--- &
\makecell{-0.03*\\{}[-0.06, -0.00]\\{}$p = 0.027$} &
--- &
\makecell{-0.06*\\{}[-0.11, -0.01]\\{}$p = 0.021$} \\

\midrule
$R^2$ & 0.323 & 0.570 & 0.137 & 0.566 & 0.157 & 0.566 & 0.274 & 0.572 \\
$N$ & 331 & 331 & 331 & 331 & 331 & 331 & 331 & 331 \\
\bottomrule
\end{tabular}
}
\vspace{0.5em}
\caption{\textbf{Longitudinal associations between sustained social chatbot engagement and subsequent well-being among continued users.} Models and estimates correspond to those reported in Table~\ref{tab:sem_persistence_combined_wdu}, with analyses restricted to participants who continued using chatbots at follow-up ($N=331$). All estimates are reported as standardized coefficients with 95\% confidence intervals and two-sided $p$-values. $^{*}p<0.05$; $^{**}p<0.01$; $^{***}p<0.001$.
}
\label{tab:sem_persistence_combined_331}
\end{table*}

%% file: tables/sem_displacement_combined_331.tex
\begin{table*}[ht]
\small
\renewcommand\cellalign{cc}
\renewcommand{\arraystretch}{1}
\setcellgapes{2pt}
\makegapedcells

\resizebox{\textwidth}{!}{%
\begin{tabular}{@{}lccccccccc@{}}
\toprule
 & \multicolumn{3}{c}{\makecell{Interaction intensity}}
 & \multicolumn{3}{c}{\makecell{Companionship (free-text)}}
 & \multicolumn{3}{c}{\makecell{Self-disclosure}} \\
\cmidrule(lr){2-4}\cmidrule(lr){5-7}\cmidrule(lr){8-10}
 & \makecell[c]{(1) \\ Interaction \\ intensity \\ T$_2$}
 & \makecell[c]{(2) \\ Time in \\ person \\ T$_{2}$} 
 & \makecell[c]{(3) \\ Well-being \\ T$_{2}$}
 & \makecell[c]{(4) \\ Companionship \\ T$_{2}$} 
 & \makecell[c]{(5) \\ Time in \\ person \\ T$_{2}$} 
 & \makecell[c]{(6) \\ Well-being \\ T$_{2}$}
 & \makecell[c]{(7) \\ Self-disclosure \\ T$_{2}$} 
 & \makecell[c]{(8) \\ Time in \\ person \\ T$_{2}$} 
 & \makecell[c]{(9) \\ Well-being \\ T$_{2}$} \\
\midrule

Interaction intensity T$_{1}$ & \makecell{0.51***\\{}[0.42, 0.60]\\{}$p < 0.001$} &
--- &
\makecell{-0.00\\{}[-0.11, 0.11]\\{}$p = 0.940$} &
--- &
--- &
--- &
--- &
--- &
--- \\

Companionship T$_{1}$ & --- &
--- &
--- &
\makecell{0.37***\\{}[0.27, 0.47]\\{}$p < 0.001$} &
--- &
\makecell{0.01\\{}[-0.06, 0.08]\\{}$p = 0.858$} &
--- &
--- &
--- \\

Self-disclosure T$_{1}$ & --- &
--- &
--- &
--- &
--- &
--- &
\makecell{0.50***\\{}[0.40, 0.59]\\{}$p < 0.001$} &
--- &
\makecell{0.01\\{}[-0.10, 0.12]\\{}$p = 0.839$} \\

Interaction intensity T$_{2}$ & --- &
\makecell{-0.16**\\{}[-0.27, -0.05]\\{}$p = 0.004$} &
\makecell{-0.11**\\{}[-0.20, -0.03]\\{}$p = 0.009$} &
--- &
--- &
--- &
--- &
--- &
--- \\

Companionship T$_{2}$ & --- &
--- &
--- &
--- &
\makecell{-0.11*\\{}[-0.21, -0.01]\\{}$p = 0.027$} &
\makecell{-0.07\\{}[-0.15, 0.00]\\{}$p = 0.057$} &
--- &
--- &
--- \\

Self-disclosure T$_{2}$ & --- &
--- &
--- &
--- &
--- &
--- &
--- &
\makecell{-0.11*\\{}[-0.22, -0.01]\\{}$p = 0.034$} &
\makecell{-0.11*\\{}[-0.21, -0.01]\\{}$p = 0.034$} \\

Time in person T$_{2}$ & --- &
--- &
\makecell{0.14***\\{}[0.06, 0.22]\\{}$p < 0.001$} &
--- &
--- &
\makecell{0.15***\\{}[0.07, 0.23]\\{}$p < 0.001$} &
--- &
--- &
\makecell{0.14***\\{}[0.07, 0.22]\\{}$p < 0.001$} \\

Well-being T$_{1}$ & --- &
--- &
\makecell{0.73***\\{}[0.65, 0.80]\\{}$p < 0.001$} &
--- &
--- &
\makecell{0.73***\\{}[0.66, 0.81]\\{}$p < 0.001$} &
--- &
--- &
\makecell{0.73***\\{}[0.66, 0.81]\\{}$p < 0.001$} \\

Chatbot tenure & \makecell{0.00\\{}[-0.09, 0.10]\\{}$p = 0.928$} &
\makecell{0.06\\{}[-0.05, 0.16]\\{}$p = 0.278$} &
\makecell{0.03\\{}[-0.04, 0.10]\\{}$p = 0.428$} &
\makecell{0.06\\{}[-0.05, 0.16]\\{}$p = 0.283$} &
\makecell{0.05\\{}[-0.06, 0.15]\\{}$p = 0.376$} &
\makecell{0.02\\{}[-0.05, 0.09]\\{}$p = 0.589$} &
\makecell{-0.02\\{}[-0.12, 0.07]\\{}$p = 0.638$} &
\makecell{0.04\\{}[-0.06, 0.14]\\{}$p = 0.462$} &
\makecell{0.02\\{}[-0.05, 0.08]\\{}$p = 0.668$} \\

Age & \makecell{0.13**\\{}[0.05, 0.21]\\{}$p = 0.002$} &
\makecell{-0.03\\{}[-0.14, 0.07]\\{}$p = 0.541$} &
\makecell{0.05\\{}[-0.03, 0.13]\\{}$p = 0.261$} &
\makecell{0.19***\\{}[0.09, 0.28]\\{}$p < 0.001$} &
\makecell{-0.05\\{}[-0.15, 0.05]\\{}$p = 0.342$} &
\makecell{0.03\\{}[-0.05, 0.11]\\{}$p = 0.438$} &
\makecell{0.08*\\{}[0.00, 0.16]\\{}$p = 0.046$} &
\makecell{-0.05\\{}[-0.16, 0.05]\\{}$p = 0.331$} &
\makecell{0.03\\{}[-0.05, 0.11]\\{}$p = 0.405$} \\

Male & \makecell{-0.11*\\{}[-0.20, -0.02]\\{}$p = 0.015$} &
\makecell{-0.28***\\{}[-0.38, -0.18]\\{}$p < 0.001$} &
\makecell{0.03\\{}[-0.05, 0.11]\\{}$p = 0.475$} &
\makecell{0.02\\{}[-0.09, 0.12]\\{}$p = 0.732$} &
\makecell{-0.25***\\{}[-0.35, -0.15]\\{}$p < 0.001$} &
\makecell{0.05\\{}[-0.03, 0.13]\\{}$p = 0.196$} &
\makecell{-0.04\\{}[-0.14, 0.05]\\{}$p = 0.380$} &
\makecell{-0.26***\\{}[-0.37, -0.16]\\{}$p < 0.001$} &
\makecell{0.04\\{}[-0.04, 0.12]\\{}$p = 0.349$} \\

Non-binary & \makecell{-0.05\\{}[-0.16, 0.06]\\{}$p = 0.362$} &
\makecell{-0.03\\{}[-0.15, 0.10]\\{}$p = 0.675$} &
\makecell{-0.00\\{}[-0.05, 0.05]\\{}$p = 0.965$} &
\makecell{-0.00\\{}[-0.11, 0.10]\\{}$p = 0.967$} &
\makecell{-0.03\\{}[-0.16, 0.10]\\{}$p = 0.672$} &
\makecell{-0.00\\{}[-0.05, 0.05]\\{}$p = 0.965$} &
\makecell{-0.05\\{}[-0.14, 0.05]\\{}$p = 0.335$} &
\makecell{-0.03\\{}[-0.15, 0.10]\\{}$p = 0.652$} &
\makecell{-0.00\\{}[-0.05, 0.04]\\{}$p = 0.872$} \\

Single & \makecell{0.07\\{}[-0.02, 0.16]\\{}$p = 0.142$} &
\makecell{-0.02\\{}[-0.13, 0.09]\\{}$p = 0.727$} &
\makecell{-0.02\\{}[-0.09, 0.06]\\{}$p = 0.631$} &
\makecell{0.05\\{}[-0.05, 0.16]\\{}$p = 0.337$} &
\makecell{-0.03\\{}[-0.13, 0.08]\\{}$p = 0.655$} &
\makecell{-0.02\\{}[-0.10, 0.05]\\{}$p = 0.543$} &
\makecell{0.03\\{}[-0.07, 0.13]\\{}$p = 0.530$} &
\makecell{-0.02\\{}[-0.13, 0.08]\\{}$p = 0.657$} &
\makecell{-0.02\\{}[-0.10, 0.05]\\{}$p = 0.568$} \\

Inverse Mills ratio & \makecell{0.04\\{}[-0.04, 0.13]\\{}$p = 0.329$} &
\makecell{0.18***\\{}[0.07, 0.28]\\{}$p < 0.001$} &
\makecell{-0.02\\{}[-0.10, 0.06]\\{}$p = 0.649$} &
\makecell{0.01\\{}[-0.09, 0.12]\\{}$p = 0.795$} &
\makecell{0.16**\\{}[0.05, 0.26]\\{}$p = 0.003$} &
\makecell{-0.04\\{}[-0.12, 0.04]\\{}$p = 0.343$} &
\makecell{0.08\\{}[-0.02, 0.18]\\{}$p = 0.112$} &
\makecell{0.17**\\{}[0.06, 0.27]\\{}$p = 0.002$} &
\makecell{-0.03\\{}[-0.11, 0.05]\\{}$p = 0.466$} \\

\midrule

Direct effect & --- &
--- &
\makecell{-0.11**\\{}[-0.20, -0.03]\\{}$p = 0.009$} &
--- &
--- &
\makecell{-0.07\\{}[-0.15, 0.00]\\{}$p = 0.057$} &
--- &
--- &
\makecell{-0.11*\\{}[-0.21, -0.01]\\{}$p = 0.034$} \\

\makecell[l]{Indirect effect via \\ $T_2$ human social \\ interaction} & --- &
--- &
\makecell{-0.02**\\{}[-0.05, -0.01]\\{}$p = 0.005$} &
--- &
--- &
\makecell{-0.02*\\{}[-0.04, -0.00]\\{}$p = 0.030$} &
--- &
--- &
\makecell{-0.02*\\{}[-0.04, -0.00]\\{}$p = 0.040$} \\

\midrule
$R^2$ & 0.323 & 0.116 & 0.576 & 0.157 & 0.104 & 0.573 & 0.274 & 0.104 & 0.579 \\
$N$ & 331 & 331 & 331 & 331 & 331 & 331 & 331 & 331 & 331 \\
\bottomrule
\end{tabular}
}
\vspace{0.5em}
\caption{\textbf{Displacement pathway between social chatbot engagement and well-being through in-person social interaction among continued users.} Models and estimates correspond to those reported in Table~\ref{tab:sem_displacement_combined_wdu}, with analyses restricted to participants who continued using chatbots at follow-up ($N=331$). All estimates are reported as standardized coefficients with 95\% confidence intervals and two-sided $p$-values. $^{*}p<0.05$; $^{**}p<0.01$; $^{***}p<0.001$.}
\label{tab:sem_displacement_combined_331}
\end{table*}

%% file: prompts/relationship_classification.tex
{\small
\begin{lstlisting}[breaklines=true]
{
    "role": "system",
    "content": """
        You are a qualitative coding assistant.

        Task: Classify a survey response about how someone views or uses chatbots. Independently decide, for EACH of the four usage types below, whether the response shows any sign of that type of use.

        Usage types (definitions):
        - Productivity: Using chatbots to obtain assistance or information.
        - Entertainment: Using chatbots for fun or to pass the time.
        - Social/Relational: The user relates to the chatbot as a personal or interpersonal counterpart, or uses it to strengthen social ties with other people. 
        - Novelty/Curiosity: Using chatbots out of curiosity or to explore their capabilities.

        Labeling rules:
        - For each type, label 1 if the response contains ANY indication of that use (even if it is not the main purpose), otherwise label 0.
        - The four types are NOT mutually exclusive: a response can be 1 on multiple types.
        - Judge each type independently.

        Examples:
        Input: 'assistant for coding and research' -> {"productivity": 1, "entertainment": 0, "social_relational": 0, "curiosity": 0}
        Input: 'I talk to it like a friend sometimes, and for fun' -> {"productivity": 0, "entertainment": 1, "social_relational": 1, "curiosity": 0}
        Input: 'helper, adviser, mentor' -> {"productivity": 1, "entertainment": 0, "social_relational": 1, "curiosity": 0}
        Input: 'roleplay partner' -> {"productivity": 0, "entertainment": 1, "social_relational": 1, "curiosity": 0}
        Input: 'storytelling and roleplaying a fictional character for my writing' -> {"productivity": 0, "entertainment": 1, "social_relational": 0, "curiosity": 0}
        Input: 'wanted to see what it could do, and it also helps me draft emails' -> {"productivity": 1, "entertainment": 0, "social_relational": 0, "curiosity": 1}
        Input: 'my companion to vent to, and it helps with homework' -> {"productivity": 1, "entertainment": 0, "social_relational": 1, "curiosity": 0}

        Reply with ONLY a JSON object with exactly these keys and integer values 0 or 1:
        {"productivity": 0, "entertainment": 0, "social_relational": 0, "curiosity": 0}
    """
}
\end{lstlisting}
}
\FloatBarrier